\documentclass[12pt,a4paper]{article}
\usepackage{amssymb}
\usepackage{sw20elba}

\input tcilatex2.5

\begin{document}

\title{\textbf{Principles of Discrete Time Mechanics:}\\
$\mathbf{V.}$\textbf{\ The Quantisation of Maxwell's Equations}}
\author{George Jaroszkiewicz and Keith Norton \\
Department of Mathematics, University of Nottingham\\
University Park, Nottingham NG7 2RD, UK}
\date{\today }
\maketitle

\begin{abstract}
\textit{Principles of discrete time mechanics are applied to the
quantisation of Maxwell's equations. Following an analysis of temporal node
and link variables, we review the classical discrete time equations in the
Coulomb and Lorentz gauges and conclude that electro-magneto duality does
not occur in pure discrete time electromagnetism. We discuss the role of
boundary conditions in our mechanics and how temporal discretisation should
influence very early universe dynamics. Quantisation of the Maxwell
potentials is approached via the discrete time Schwinger action principle
and the Faddeev-Popov path integral. We demonstrate complete agreement in
the case of the Coulomb gauge, obtaining the vacuum functional and the
discrete time field commutators in that gauge. Finally, we use the
Faddeev-Popov method to construct the discrete time analogues of the photon
propagator in the Landau and Feynman gauges, which casts light on the break
with relativity and possible discrete time analogues of the metric tensor. }
\end{abstract}

Throughout this paper the acronym \textit{CT} refers to \textit{continuous
time}, whereas \textit{DT} refers to \textit{discrete time}. Readers
familiar with the principles and methodology discussed in the earlier papers
of this series may skip the introduction, but are advised that the general
notation has been improved and is discussed in section 2.

\section{Introduction}

The development of quantum field theory in the second quarter of the
twentieth century was accompanied by speculation about the microscopic
nature of time and space, motivated by the spectacle of quantised fields
dynamically evolving over a classical space-time in no way different to the
bland Riemannian space-time continuum used by Einstein in general
relativity. Various problems in quantum field theory such as the divergences
in the renormalisation programme and ambiguities in operator products were
believed to be associated in some way with the microscopic description of
space-time, but relatively little was done to investigate the issue in any
depth. Although there were occasional attempts to investigate alternative
mathematical descriptions of space-time, such as the notable work of Snyder 
\cite{SNYDER-47A,SNYDER-47B}, it was more usual to circumvent difficulties
with ad hoc procedures such as point splitting of operator products with no
modification of the underlying space-time, or by the use of space-time
lattices, which were always assumed to be an approximation to the continuum.
It was only with the advent of the space-time foam approach to quantum
gravity and the reinterpretation of superstring theory in the nineteen
eighties that it became generally acceptable to talk about the microscopic
nature of space-time as more than likely quite different to the continuum
normally assumed in field theory. It is in this context that our work should
be seen. \nocite{JAROSZKIEWICZ-97A}\nocite{JAROSZKIEWICZ-97B} \nocite
{JAROSZKIEWICZ-97C}\nocite{JAROSZKIEWICZ-97D}

In the current series of paper on \textit{DT} mechanics [3-6] we investigate
the consequences of taking literally the hypothesis that time is discrete on
an incredibly small scale. Our original motivation is discussed in the first
paper of this series \cite{JAROSZKIEWICZ-97A}. Compared with quantum gravity
and superstring theory, ours is a very modest and limited step. However, it
turns out to have enormous consequences, principally conceptual, altering
virtually all aspects of the laws of mechanics and how we view space-time.
Along the way a number of sacred cows have to be sacrificed. For instance,
without continuity with respect to time there is no differentiation with
respect to time. Therefore we are forced into the construction of a
mechanics without velocities. This immediately raises the question of what
replaces Lagrangians, which are normally functions of dynamical variables
and their temporal derivatives. Since we have no velocities in our theory,
we cannot construct conjugate momenta in the traditional way, as these are
defined as derivatives of Lagrangians with respect to velocities. So we
appear not to have a phase space, and consequently we do not have a
Hamiltonian formulation or Poisson brackets in the normal sense of the word.
This goes hand in hand with the lack of continuous translations in time and
with the absence of a generator of such transformations. This raises
questions about quantisation, but we have shown in earlier papers of this
series that these can answered. The price we pay for this is that we end up
no longer doing exactly what we were doing before, amounting to a paradigm
shift in the language of Thomas Kuhn.

We regard it as an important principle that we do not simply modify \textit{%
CT} equations of motion by replacing temporal derivatives with ad hoc
differences. Although that works in some situations in Newtonian mechanics,
it becomes more subtle in the presence of gauge invariance. Our approach is
to start from the beginning, rewriting the action integral as an action sum
and developing the consequences rigorously from there. In this series of
papers we have not confined our interest to particular models which happen
to be amenable to temporal discretisation. The microscopic nature of space
and time will affect all dynamical processes. Our interest has been in the
subject of \textit{DT }mechanics as a whole, in both its classical and
quantised forms, applied to point and field systems.

An important consequence of temporal discretisation is that it rejects the
notion that space and time form a four dimensional continuum. This notion
has been successfully exploited in the special and general theories of
relativity throughout this century and is a cornerstone of the theories of
quantum gravity and superstring theory (which paradoxically eventually
undermine this very idea). We appear to have taken a step back towards the
separation of absolute space and time in Newtonian mechanics. Discretising
time also raises questions any relativist would ask, which are: \textit{in
which inertial frame is time discrete and what dictates this choice?} Our
discrete time mechanics is not Lorentz covariant, and we have to address
this issue as well as others.

We believe that there are satisfactory answers to these particular questions
which accord with modern cosmology. Consider a \textit{CT}
Freidmann-Robertson-Walker space-time, i.e., one for which \textit{CT}
co-ordinates can be chosen so that the Riemannian pseudo-metric distance
rule takes the standard form 
\begin{equation}
ds^2=dt^2-R\left( t\right) ^2d\sigma ^2.  \label{FRW}
\end{equation}
Here $R\left( t\right) $ is a function of the co-moving time $t$ only and $%
d\sigma $ is the distance element of some spherical, flat, or hyperbolic $3-$%
space. Given that the gross space-time structure of the expanding universe
is well represented by such a choice of co-ordinates and by such a distance
rule, consider now the cosmic microwave background radiation $\left( \text{%
\textit{CMB}}\right) $ field discovered by Penzias and Wilson in 1964 \cite
{PENZIAS-69}. It has been pointed out by a number of authors \cite
{STAPP-93,HAWLEY_&_HOLCOMB}, that, contrary to the principles of relativity,
this radiation field can be used to define a local absolute inertial frame $%
F_P$ at each point $P$ in space-time. Such a frame is unique up to Euclidean
transformations such as rotations. For an observer at $P$ instantaneously at
rest relative to $F_P$, the \textit{CMB} radiation field will appear
isotropic to a very high degree. That this is physically meaningful is borne
out by the empirical observation that the earth appears to be moving at a
speed of about $500-600$ km/sec relative the local $F_P$ frame, a phenomenon
called the \textit{dipole effect }\cite{SMOOT}\textit{.} It is somewhat
ironic that long after the Michelson-Morley experiment failed to detect an 
\emph{aether }carrying radiation (and thereby supporting the principle of
special relativity), Penzias and Wilson discovered a plenum consisting
entirely of radiation which may be used to define absolute local inertial
frames.

Our thinking is as follows. Suppose we take seriously the hypothesis that
time is really discrete. Our fundamental criterion is that any dynamics
based on this idea should not make predictions at odds with scientifically
determined (i.e. empirical) facts. We shall call this the \textit{empirical
principle. }It has been part of our programme to determine where if anywhere
temporal discretisation actually clashes with this principle. So far, we
have not been able to rule \textit{DT} mechanics out on this basis.

Always mindful of this principle, we should now carefully sift out and
identify those additional concepts and ideas which, despite our traditional
training and inclination, are really no more than contemporary belief
structures which happen to be compatible with the empirical principle. We
shall call such ideas \textit{idealisations. }We should feel free to
dispense with idealisations if absolutely necessary, provided we do not
clash with the empirical principle. It is particularly important to identify
idealisations which are currently popular because of their elegance and
mathematical content; these are usually the hardest to remove, because
mathematical elegance is frequently taken as a principle in physics.

A good example of such an idealisation is the Poincar\'{e} group. This
particular structure cannot occur in our theory because we do not have the
freedom to make continuous displacements in time; this property emerges only
in the continuous time limit. What replaces the Poincar\'{e} group in 
\textit{DT} mechanics is still under investigation. What can be said however
is that any \textit{DT }analogue will almost certainly be more complicated
mathematically. The absence of the Poincar\'{e} group however does not
unduly worry us here and should not be used as a criticism, because our
mechanics can still satisfy the empirical principle for the following
reason. In our \textit{DT} mechanics there is a fundamental interval of time 
$T$ and generally, we can show agreement with \textit{CT }theory certainly
at the $O\left( T^0\right) $ level$.$ Moreover, disagreement invariably
occurs at the $O\left( T^2\right) $ level. Since we imagine $T$ is of the
order of the Planck time $T_P\equiv \sqrt{\hbar G/c^5}\simeq 5\times
10^{-44}\sec ,$ our theory should be good for the current level of
experimental accuracy. As yet we have no explanation of the value of $T$,
but then we cannot explain the value of $c,$ $\hslash ,G$ or the electric
charge either.

Continuing this line of thought, we argue that temporal discreteness would
have influenced both the dynamical origin of the universe and its subsequent
evolution. The \textit{Planck epoch} is a term used to denote the interval
from the origin of time to $T_P.$ By the end of that interval, it is
generally believed that gravity had decoupled from the other interactions
and space started to expand in a pre-inflationary context. It is during the
Planck epoch that \textit{CT} field theories are generally considered to be
either invalid or seriously incomplete, and it is possible that some version
of \textit{DT }mechanics (though not necessarily the one we are considering
here) is the appropriate theory to describe very early universe dynamics. If
true, then a generally covariant description in the fashion of general
relativity might not be at all appropriate during this epoch, with the best
description perhaps involving some preferred frame of reference.

It may be the case that after the Planck time and before false vacuum
inflation, different regions of the Universe had different local temporal
discretisation frames, randomly distributed, analogous to pre-inflationary
early universe monopoles or ferromagnetic domains in a solid. If so, we
would argue that in much the same way as the \textit{monopole problem} was
removed by inflation, only one actual temporal discretisation frame would
survive inflation into our local universe. All the others would be beyond
the event horizon.

The discretisation frame holding in our visible universe would most likely
be linked to the dynamical processes involved with the phase transition from
the false vacuum to radiation and matter, and we would expect it to leave a
signature. This signature would be the existence of the co-moving frame used
in the FRW metric (\ref{FRW}), and subsequently, the isotropy frame of the 
\textit{CMB.} We propose that the discrete time parameter discussed in this
paper be identified as the discretised version of the local co-moving time
(assumed to be the coordinate time in our local $F_P)$ in our neighbourhood.
We are assuming here that the \textit{CMB} isotropy frame and the cosmic
matter rest frame coincide locally \cite{HAWLEY_&_HOLCOMB}.

Now over laboratory scales associated with particle scattering experiments,
we could ignore the local aspect of this discretisation and regard a global
discrete time frame as a very good approximation, in much the same way that
special relativity is a good approximation to general relativity in the
laboratory. Hence we end up with the equivalent of a unique discretisation
of Minkowski space-time in our neighbourhood.

A \textit{DT} analogue of general relativity awaits investigation; it will
require abandoning the principle of general covariance on a microscopic
level, with a generally covariant description emerging only in the \textit{CT%
} limit. It is possible that our fundamental interval $T$ is itself
determined dynamically by the local matter densities and by the dynamics.
There is a precedent for this idea. In 1983 Lee published a study of
discrete time field theory \cite{LEE-83}, where his interval of time $T_n$
varied and was a dynamical variable coupled to matter. In our earlier work
and in this paper, time is discrete but otherwise is as passive as time in 
\textit{CT} relativistic field theory. We work with a fixed $T$ throughout.

In this paper we develop further the discrete time Maxwell's equations first
discussed in \cite{JAROSZKIEWICZ-97A,JAROSZKIEWICZ-97B}. In the next section
we review our notation, which has been somewhat overhauled and compactified
compared to earlier papers in the series, but with no change in content.
Following that, we review the \textit{DT }mechanics formalism, extending the
discussion to include the dynamical variables associated with temporal
links. Our view of discrete time mechanics has evolved from our original
picture of dynamical variables changing over successive instants of time (%
\textit{nodes}) only to one where there are dynamical variables on the nodes
and on the intervals or \textit{links }between the nodes.\textit{\ }A gauge
theory such as Maxwell's equations (and undoubtedly gravitation)\textit{\ }%
involves a dynamical interplay between node variables and link variables.

Then we review the \textit{DT }Maxwell's equations in the Lorentz and
Coulomb gauges. Working in the former gauge presents a particularly
interesting challenge in \textit{DT }mechanics because it is a relic of the
Lorentz symmetry which \textit{DT} mechanics undermines. At first sight is
appears less natural than the Coulomb gauge defined in the local absolute
rest frame in which time is discretised. We recall that the use of Lorentz
gauges and attempts to maintain manifest Lorentz covariance in \textit{CT }%
quantum\textit{\ }electrodynamics leads to problems with state vectors,
requiring the Gupta-Bleuler formulation or equivalent technology.

We then consider the quantisation of the free Maxwell fields. Using the 
\textit{DT }Schwinger action principle which was successfully used
previously for the \textit{DT} quantisation of the scalar \cite
{JAROSZKIEWICZ-97C} and Dirac fields \cite{JAROSZKIEWICZ-97D}, we find the
vacuum functional and the free field commutators in the Coulomb gauge.
Although we do not have a Hamiltonian formulation in our theory, we do have
gauge symmetry, and we may use the Faddeev-Popov approach to the
quantisation of gauge fields in our theory. We find that the Faddeev-Popov
vacuum functional for the \textit{DT} free electromagnetic theory in the 
\textit{DT} Coulomb gauge coincides precisely with the vacuum functional
obtained by solving the \textit{DT }Schwinger functional differential
equations in the same gauge.

Finally, and this was found to be the hardest task, we construct the \textit{%
DT} analogues of the photon propagator in the Landau and Feynman gauges, in
preparation for future applications to QED. We find that there is an
interesting breakdown of what happens in \textit{CT} theory, with \textit{DT}
analogues of the metric tensor making appearances. These differ in
interesting ways, depending on whether we are in space time or in momentum
space, and indicate that a reformulation of general relativity via \textit{DT%
} mechanics will be quite instructive.

\section{Notation and conventions}

In addition to the natural unit system where $c=\hbar =1,$ we use the
following conventions throughout this paper. Given $T$ is the fundamental
time step in \textit{DT} mechanics then an event in our space-time has
natural coordinates $x\equiv \left( n,\mathbf{x}\right) $, where $n$ is an
integer. The \textit{CT} limit 
\begin{equation}
T\rightarrow 0,\,n\rightarrow \infty ,\;\;\;nT=t\;\;\;(\,t\;\text{fixed})
\label{CTL}
\end{equation}
then corresponds to the\textbf{\ }\textit{CT} coordinate time $t$. Our
coordinates are natural in the following sense. We work in an inertial frame
of reference such that the \textit{CMB} dipole anisotropy is absent locally 
\cite{SMOOT}. In this frame we fix cartesian spatial axes with coordinates $%
\mathbf{x}\equiv \left( x^1,x^2,x^3\right) $ such that the distant stars do
not appear to rotate relative to such axes. Finally we choose an instant of
discrete time as our temporal origin of coordinates and count fundamental
intervals of $T$ forwards and backwards, giving us the integer coordinate $n$
referred to above. Our discussion in this paper does not include any
cosmological aspects such as Hubble expansion, other than using the \textit{%
CMB} to determine our frame of reference. We shall be concerned with local
coordinates relevant to particle physics in the present epoch.

We denote integration over space-time by the sum/integral symbol 
\begin{equation}
\Sigma \!\!\!\!\!\!\int_x\equiv T\sum_{n=-\infty }^\infty \int d^3\mathbf{x.}
\end{equation}
Given a variable $f_x\equiv f_n\left( \mathbf{x}\right) $ indexed by a
discrete temporal index $n$ and by a continuous spatial index $\mathbf{x}$
we define the \textit{DT }fourier transform $\tilde{f}_p$ of $f_x$ by 
\begin{equation}
\tilde{f}_p\equiv \tilde{f}\left( z,\mathbf{p}\right) \equiv \Sigma
\!\!\!\!\!\!\int_xe_{px}f_x,
\end{equation}
where 
\begin{equation}
e_{px}\equiv z^ne^{-i\mathbf{p\cdot x}}
\end{equation}
and $z$ is complex and non-zero. We shall assume that such a transform
exists and defines a function analytic in some annular region in the complex-%
$z$ plane centred on the origin and including the unit circle. This of
course puts restrictions on the sequence defined by $f_n$. Although $\mathbf{%
p}$ is the direct equivalent of the spatial components of the momentum space
four-vector of \textit{CT} mechanics, the analogue of the time component $%
p^0 $ of the momentum space four vector is \textit{not} $z.$ We shall be
interested in taking $z$ on the unit circle, and then it is the principal
argument of $z$ which is related to $p^0.$ We shall use the symbol $p$ to
denote the set $p\equiv \left( z,\mathbf{p}\right) .$

If the \textit{DT} fourier transform exists and has a Laurent expansion in
some annular region centred on the origin in the complex $z-$plane and
containing the unit circle, then we can construct the inverse transform.
This is given by 
\begin{equation}
f_x=\oint_p\bar{e}_{px}\tilde{f}_p,  \label{inverse}
\end{equation}
where 
\begin{equation}
\oint_p\equiv \frac 1{2\pi iT}\oint \frac{dz}z\int \frac{d^3\mathbf{p}}{%
\left( 2\pi \right) ^3}
\end{equation}
with the $z-$integral being over the unit circle in the complex plane taken
in the anticlockwise sense and 
\begin{equation}
\bar{e}_{px}\equiv z^{-n}e^{i\mathbf{p\cdot x}}.
\end{equation}

We could attempt to \textit{define }our dynamics in the $\left( z,\mathbf{p}%
\right) $ transform space, restricting our dynamical variables to be
functions in the complex $z-$ plane for which the inverse transform $\left( 
\ref{inverse}\right) $ exists. No new dynamical content should emerge from
this approach, but it may be mathematically more secure. However, we will
normally assume that coming first from the space-time direction is valid.
That after all is the conventional way to define field theories.

The analogues of the identity operators (delta functions) in our formalism
are defined as follows. If $\delta _n$ is the \textit{DT }Kronecker delta,
satisfying 
\begin{eqnarray}
\delta _n &=&1,\;\;\;\;n=0,  \nonumber \\
&=&0,\;\;\;\;n\neq 0
\end{eqnarray}
where $n$ is an integer, then we define the four dimensional \textit{DT }%
Dirac delta $\delta _x$ by 
\begin{equation}
\delta _{x-y}\equiv \frac{\delta _{n-m}}T\delta ^3\left( \mathbf{x-y}\right)
,\;\;\;\;\;T>0,
\end{equation}
where $y\equiv \left( m,\mathbf{y}\right) .$ Then 
\begin{equation}
\Sigma \!\!\!\!\!\!\int_xf_x\delta _{x-y}=f_y.
\end{equation}
An integral representation of $\delta _{x-y}$ is 
\begin{equation}
\delta _{x-y}=\oint_pe_{px}\bar{e}_{py}.
\end{equation}
The corresponding operator $\tilde{\delta}_{p-q}$ in \textit{DT }fourier
transform space satisfies the relation 
\begin{equation}
\oint_p\tilde{f}_p\tilde{\delta}_{p-q}=\tilde{f}_q,\;\;\;\;\;\;\;\;\;q\equiv
\left( u,\mathbf{q}\right) ,
\end{equation}
where $u$ is complex. A \textit{DT }sum/integral representation of $\tilde{%
\delta}_{p-q}$ is given by 
\begin{equation}
\tilde{\delta}_{p-q}=\Sigma \!\!\!\!\!\!\int_x\bar{e}_{px}e_{qx}.
\end{equation}

The classical step operator $U_{n\text{ }}$acting on any temporally indexed
variable $f_{n\text{ }}$ is defined by 
\begin{equation}
U_nf_n\equiv f_{n+1},\;\;\;U_n^{-1}f_n\equiv f_{n-1},
\end{equation}
with powers of $U$ defined in the obvious way, viz 
\begin{equation}
U_n^af_n=f_{n+a},
\end{equation}
where $a$ is an integer. We note that 
\begin{equation}
U_nf_m=f_m,\;\;\;\;m\neq n.
\end{equation}

Throughout our mechanics we shall deal with operators and equations defined
via linear combinations of the step operators. An operator of the form 
\begin{equation}
P\left( U_n\right) \equiv c_0U_n^a+c_1U_n^{a+1}+...+c_rU_n^{a+r}
\end{equation}
will be called an $r^{th}$ \textit{order }operator. Then an equation
involving an $r^{th}$ order operator will be called $r^{th}$ order.
Important first order operators are the forwards and backwards differences
defined by 
\begin{equation}
\Delta _n^{+}\equiv U_n-1,\;\;\;\;\;\Delta _n^{-}\equiv 1-U_n^{-1}
\end{equation}
respectively, and from them we define the second order symmetric difference 
\begin{equation}
\Delta _n\equiv \frac{_1}{^2}\left( \Delta _n^{+}+\Delta _n^{-}\right) =%
\frac{_1}{^2}\left( U_n-U_n^{-1}\right) .
\end{equation}
First temporal derivatives are invariably replaced by one of three possible
operators, defined by 
\begin{equation}
D_n^{+}\equiv \frac{\Delta _n^{+}}T,\;\;D_n^{-}\equiv \frac{\Delta _n^{-}}%
T,\;\;\;D_n\equiv \frac{\Delta _n}T,
\end{equation}
whereas the second derivative is invariably replaced by the operator 
\begin{equation}
D_n^2\equiv D_n^{+}D_n^{-}=D_n^{-}D_n^{+}=\frac{U_n-2+U_n^{-1}}{T^2}.
\end{equation}
It is a feature of our \textit{DT }mechanics that the formalism will tell us
which of the above difference operators we need to use in a given context.
For example, our investigations into the \textit{DT} Schr\"{o}dinger
equation \cite{JAROSZKIEWICZ-97B} and the Dirac equation \cite
{JAROSZKIEWICZ-97C} show that in those equations we have to replace $%
\partial /\partial t$ by the second order symmetric operator $D_n$. This has
important consequences as far as the solutions of the equations are
concerned, because the second order symmetric difference leads to a \textit{%
DT} equation of motion which acts like a second order \textit{CT }equation
of motion, rather than a first order equation of motion, and this generates
the \textit{oscillon} solutions discussed in \cite{JAROSZKIEWICZ-97B} and 
\cite{JAROSZKIEWICZ-97C}. Fortunately, we found that in the second quantised
theory, these oscillons correspond to states with unphysical norm, and so
are not observable asymptotically as ordinary particles. Their role as
virtual particles in \textit{QED }will be considered in the next paper of
this series.

In our mechanics, \textit{time reversal} amounts to the interchange $%
U_n\leftrightarrow U_n^{-1}.$ We shall encounter various difference
operators which have the symmetry property that they are invariant to time
reversal. We shall refer to such operators as $T-$\textit{symmetric. }They
are important and useful to us, and in \textit{DT }fourier transform space
they are real functions of $z$ provided $z$ is on the unit circle (we note
that $z^{*}=z^{-1}$ holds only on the unit circle). With this definition, we
see $D_n^2$ is a second order $T$-symmetric operator. Another important
second order $T$-symmetric operator which occurs throughout our mechanics is
given by 
\begin{equation}
S_n\equiv \frac{_1}{^6}\left( U_n+4+U_n^{-1}\right) .
\end{equation}
The factor of $1/6$ and the $4$ can readily be understood in terms of our
virtual path procedure, discussed below. In the \textit{CT }limit $\left( 
\ref{CTL}\right) ,$ if it exists, these operators can be replaced by 
\begin{eqnarray}
D_n^{+} &\rightarrow &\partial _t,\;\;D_n^{-}\rightarrow \partial
_t,\;\;\;D_n\rightarrow \partial t, \\
D_n^2 &\rightarrow &\partial _t^2,\;\;\;\;S_n\rightarrow 1.
\end{eqnarray}
There is no trace of $S_n$ in \textit{CT} theory, but it plagues \textit{DT }%
mechanics\textit{, }occurring in unpredictable places and introducing
temporal nonlocality in unexpected places. It makes the rewriting of \textit{%
CT} mechanics into a \textit{DT} form far from easy, particularly in the
case of Maxwell's equations. This nonlocality also enters at the level of
the metric tensor, and suggests that an attempt to formulate general
relativity into a \textit{DT }framework will require thinking of the metric
tensor as a non-local in time operator (and hence as a more dynamical
object), rather than as a set of local functions forming the components of a
rather bland second rank tensor. Another important operator is the \textit{DT%
} d'Alembertian, which turns out to be the second order $T$-symmetric
difference-differential operator 
\begin{equation}
\Box _x\equiv D_n^2-S_n\nabla _{\mathbf{x}}^2,
\end{equation}
which appeared in our studies of the \textit{DT} Klein Gordon equation \cite
{JAROSZKIEWICZ-97B,JAROSZKIEWICZ-97C} and the \textit{DT} Dirac equation 
\cite{JAROSZKIEWICZ-97D}. This operator is also important in our formulation
of Maxwell's equations.

We note the useful result 
\begin{equation}
\Sigma \!\!\!\!\!\!\int_xf_x\overrightarrow{P\left( U_n\right) }g_x=\Sigma
\!\!\!\!\!\!\int_xf_x\overleftarrow{P\left( U_n^{-1}\right) }g_x.
\end{equation}
{}From this we see that for $T$-symmetric operators, we may write 
\begin{equation}
\Sigma \!\!\!\!\!\!\int_xf_x\overrightarrow{P\left( U_n\right) }g_x=\Sigma
\!\!\!\!\!\!\int_xf_x\overleftarrow{P\left( U_n^{}\right) }g_x.
\end{equation}
The \textit{DT} fourier transform gives the following useful result: 
\begin{equation}
\Sigma \!\!\!\!\!\!\int_xe_{px}P\left( U_n\right) f_x=P\left( z^{-1}\right) 
\tilde{f}_p,\;\;\;\;\;z\neq 0.
\end{equation}

We may use the above to find the \textit{DT }Feynman propagator $\Delta
_{Fx} $, which satisfies the equation \cite{JAROSZKIEWICZ-97B} 
\begin{equation}
\left( \square _x+m^2S_n\right) \Delta _{Fx}=-\delta _x.
\end{equation}
Taking the \textit{DT} fourier transform of this equation we find 
\begin{equation}
\left( p^2-m^2Sz\right) \tilde{\Delta}_{Fp}=1,
\end{equation}
where we define 
\begin{equation}
p^2\equiv -D^2z-Sz\mathbf{p}.\mathbf{p},
\end{equation}
with 
\begin{equation}
D^2z\equiv \frac{z-2+z^{-1}}{T^2},\;\;\;\;\;\;Sz\equiv \frac{z+4+z^{-1}}6.
\end{equation}
A solution of interest in particle theory is 
\begin{equation}
\Delta _{Fx}=\oint_p\bar{e}_{px}\frac 1{p^2-m^2Sz+i\epsilon },  \label{form}
\end{equation}
where we choose a \textit{DT} analogue of the Feynman $+i\epsilon $
prescription. The singularity structure in the complex $z$-plane of the
integrand in the above is particularly interesting. First, we can prove that
the equation 
\begin{equation}
p^2-m^2Sz+i\epsilon =0  \label{root}
\end{equation}
has no solution on the unit circle in the complex $z-$plane for any value of
the linear momentum $\mathbf{p}$. To prove this, write $z=\exp \left(
i\theta \right) .$ Then $\left( \text{\ref{root}}\right) $ becomes 
\begin{equation}
\frac{2\left( \cos \theta -1\right) }{T^2}-\frac{\left( \cos \theta
+2\right) }3\left( \mathbf{p\cdot p+}m^2\right) +i\epsilon =0,
\end{equation}
which has no solution for real $\theta $ if $\epsilon >0.$ This result is
important because it means we have a fully closed contour of integration
over the unit circle in the complex $z-$plane, requiring no principal value
discussion.

Next, we note that $\left( \text{\ref{root}}\right) $ is a quadratic in $z$,
with roots $z_1$, $z_2$ satisfying the relation 
\begin{equation}
z_1z_2=1.
\end{equation}
Hence we deduce that the denominator in $\left( \text{\ref{form}}\right) $
contributes one simple pole inside the unit circle and one outside. By
looking carefully at the location of these poles as the spatial momentum $%
\mathbf{p}$ varies, we see that two distinct patterns of behaviour emerge 
\cite{JAROSZKIEWICZ-97A}. For momentum in the \textit{elliptic regime, }%
corresponding to momenta bounded by $T|\mathbf{p}|<\sqrt{12},$ the simple
pole inside the unit circle is just inside and gives the equivalent of a
trigonometric solution when we use the calculus of residues. For momentum in
the \textit{hyperbolic regime}, on the other hand, given by $T|\mathbf{p}|>%
\sqrt{12},$ the simple pole interior to the unit circle starts to move
towards $z=0$, giving a damped exponential solution when we use the calculus
of residues.

Finally, we note that if we had taken a slightly different prescription, viz 
\begin{equation}
\Delta _{Fx}\equiv \oint_p\bar{e}_{px}^{}\frac 1{p^2-\left( m^2-i\epsilon
\right) Sz}
\end{equation}
it is not hard to see that our conclusions would be exactly the same.

\section{Link and node variables in \textit{DT }mechanics}

The discretisation of time requires us to change the way we think about
dynamical variables. Given a lattice structure to time, we can identify two
distinct geometrical components. These are the \textit{node} sites, which
correspond to events at times $nT$, where $n$ is an integer, and the \textit{%
links} between these nodes. For a one dimensional lattice, links and nodes
are mathematically dual, but in our theory this mathematical duality does
not carry over into the physics. Some of our dynamical variables are defined
at nodes whereas others are defined at sites, and generally this occurs in
such a way that interchanging links and nodes is not a symmetry of the
theory. In particular, matter fields corresponding to massive particles are
defined on nodes only. We have found that electric fields occur as link
variables whereas magnetic fields are node variables, so that
electro-magnetic duality does not appear to occur here. This seems a natural
way to explain the absence of Maxwellian magnetic monopoles.

We show now that the dynamical rules for link variables are similar to but
not identical to those for node variables. First, let a generic node field
dynamical variable be denoted by $A_n^\alpha \left( \mathbf{x}\right) $,
where $n$ is the time, $\mathbf{x}$ is the spatial position, and $\alpha $
is some extra label such as a vector, spinorial or group index. Likewise,
denote a typical link variable by $\phi _n^\beta \left( \mathbf{x}\right) $.
As discussed in \cite{JAROSZKIEWICZ-97A,JAROSZKIEWICZ-97B} we base our
dynamics on the \textit{system function }$F^n\equiv \int d\mathbf{x\,}%
\mathcal{F}^n$, where the \textit{system function density} $\mathcal{F}^n$
has the form 
\begin{equation}
\mathcal{F}^n\equiv \mathcal{F}\left( A_n,A_{n+1},\phi _n,\nabla A_n,\nabla
A_{n+1},\nabla \phi _n\right) ,
\end{equation}
where we have suppressed the spatial coordinates and field labels. This is
the discrete time analogue of a Lagrangian of the standard form 
\begin{equation}
L\equiv \int d\mathbf{x}\mathcal{L}\left( \varphi ,\nabla \varphi ,\dot{%
\varphi}\right)
\end{equation}
and is equally generic, in that all of the system functions we need to use
are of this form. We note that link and node variables are treated
differently right at this point, in that the system function is first order
in the node variables but zeroth order in the link variables. This
emphasises further that as far as dynamics is concerned, link and node
variables are not dual.

In \textit{DT} mechanics the action integral becomes an \textit{action sum}.
The action sum from initial time $MT$ to final time $NT>MT$ is given by 
\begin{equation}
A^{NM}\equiv T\sum_{n=M}^{N-1}F^n
\end{equation}
and the equations of motion are obtained by Cadzow's action principle \cite
{CADZOW-70} 
\begin{equation}
\delta A^{NM}\stackunder{c}{=}0,
\end{equation}
for suitable variations of the fields. Here and elsewhere we shall use the
symbol $\stackunder{c}{=}$ to denote an equality holding by virtue of the
equations of motion. For an arbitrary variation: 
\begin{eqnarray}
A_n^\alpha &\rightarrow &A_n^\alpha +\delta A_n^\alpha ,\;\;\;\;M\leq n\leq N
\nonumber \\
\phi _n^\beta &\rightarrow &\phi _n^\beta +\delta \phi _n^\beta
,\;\;\;\;\;M\leq n<N
\end{eqnarray}
we find 
\begin{eqnarray}
\delta A^{NM} &=&T\int d^3\mathbf{x}\left\{ \delta A_M^\alpha \left( \mathbf{%
x}\right) \frac \delta {\delta A_M^\alpha \left( \mathbf{x}\right)
}F^M+\sum_{n=M+1}^{N-1}\delta A_n^\alpha \left( \mathbf{x}\right) \frac
\delta {\delta A_n^\alpha \left( \mathbf{x}\right) }F^n\right.  \nonumber \\
&&\;\;\;\;\;+\delta A_N^\alpha \left( \mathbf{x}\right) \frac \delta {\delta
A_N^\alpha \left( \mathbf{x}\right) }F^{N-1}+\sum_{n=M+1}^{N-1}\delta
A_n^\alpha \left( \mathbf{x}\right) \frac \delta {\delta A_n^\alpha \left( 
\mathbf{x}\right) }F^{n-1} \\
&&\;\;\;\;\;\;\;\;\;\;\;\;\;\;\;\;\;\;\;\;\;\;\;\;\;\;\;\left.
+\sum_{n=M}^{N-1}\delta \phi _n^\alpha \left( \mathbf{x}\right) \frac \delta
{\delta \phi _n^\alpha \left( \mathbf{x}\right) }F^n\right\} ,  \nonumber
\end{eqnarray}
where we use the summation convention for the field labels $\alpha $ only.
For fixed end-point, but otherwise arbitrary variations, viz. 
\begin{equation}
\delta A_M^\alpha \left( \mathbf{x}\right) =\delta A_N^\alpha \left( \mathbf{%
x}\right) =0
\end{equation}
we find the functional derivative equations of motion 
\begin{eqnarray}
\frac \delta {\delta A_n^\alpha \left( \mathbf{x}\right) }\left\{
F^n+F^{n-1}\right\} \stackunder{c}{=}0, &&\;\;\;\;\;M<n<N  \label{Cad} \\
\frac \delta {\delta \phi _n^\alpha \left( \mathbf{x}\right) }F^n\stackunder{%
c}{=}0, &&\;\;\;\;\;M\leq n<n,  \label{Cad2}
\end{eqnarray}
which reduce to 
\begin{eqnarray}
&&\frac \partial {\partial A_n^\alpha \left( \mathbf{x}\right) }\left\{ 
\mathcal{F}^n+\mathcal{F}^{n-1}\right\} \stackunder{c}{=}\nabla \cdot \frac
\partial {\partial \nabla A_n^\alpha \left( \mathbf{x}\right) }\left\{ 
\mathcal{F}^n+\mathcal{F}^{n-1}\right\} ,\;\;\;M<n<N  \label{eq1} \\
&&\;\;\;\;\;\;\;\;\;\;\;\;\;\;\;\;\frac \partial {\partial \phi _n^\beta
\left( \mathbf{x}\right) }\mathcal{F}^n\stackunder{c}{=}\nabla \cdot \frac
\partial {\partial \nabla \phi _n^\beta \left( \mathbf{x}\right) }\mathcal{F}%
^n,\;\;\;\;\;\;\;\;\;\;\;\;\;\;M\leq n<N.  \label{eq2}
\end{eqnarray}

We see here once more the essential difference between the node field
dynamics and that of the link fields. The former involve genuine second
order difference equations of motion $\left( \ref{eq1}\right) $
corresponding to second order temporal derivatives in \textit{CT} theory.
For the links however, equations $\left( \ref{eq2}\right) $ are at most
first order difference equations, and such equations are analogous to
constraint equations in \textit{CT} mechanics rather than equations of
motion. All of this is intimately tied in with the boundary conditions
required to solve these equations. \textit{CT} theories with constraints and
gauge symmetries will have \textit{DT} analogues where both sort of
equations occur. The electromagnetic equations discussed in this paper give
a basic example of these ideas.

\section{Classical electromagnetism}

\subsection{The charge free Maxwell equations}

Our discrete time formulation of the charge free Maxwell's equations starts
with the \textit{CT }electromagnetic potentials $A^\mu \equiv \left( \phi ,%
\mathbf{A}\right) $ which are used to construct the physical electric and
magnetic fields $\mathbf{E}$ and $\mathbf{B}$. A clear distinction has to be
made here between the nature of the electric scalar potential $\phi $ and
the magnetic vector potential $\mathbf{A}$ in \textit{DT} mechanics. The
former is associated with the \textit{temporal interval }or\textit{\ link }%
connecting times $t_n\equiv nT$ and $t_{n+1},$ whereas the latter is
associated with nodes, or times $t_n$ themselves. This distinction also
manifests itself in the difference between the physical electric field $%
\mathbf{E}$ and the magnetic field $\mathbf{B}$, which are likewise
associated with temporal links and nodes respectively. The electric scalar
potential associated with the link connecting time $t_n$ and $t_{n+1}$ at
spatial position $\mathbf{x}$ will be denoted by the symbol $\phi _n\left( 
\mathbf{x}\right) ,$ rather than by (say) $\phi _{n+\frac 12}\left( \mathbf{x%
}\right) .$ Although our notation suggests a bias towards $t_n$ at the
expense of $t_{n+1},$ this is not really the case.

One of the principles we have applied to \textit{DT} mechanics is \textit{%
minimality}. By this we mean our belief that if we are really working at an
irreducibly fundamental level, then the dynamics should be as basic and
uncomplicated as possible. Now as discussed in $\cite{JAROSZKIEWICZ-97A}$ a
system function is more like a Hamilton's principle function than a
Lagrangian, being a function of the dynamical degrees of freedom at the end
points of a fundamental interval of time $T$. In principle, we cannot probe
below this level in \textit{DT }mechanics. We could simply postulate a
system function, but a better approach would be to take some \textit{CT}
theory such as Maxwell's electromagnetism and construct a suitable system
function directly from its \textit{CT }Lagrangian\textit{. }If we did not
use the\textit{\ CT} Lagrangian as a guide, then we could not hope to guess
an appropriate system function.

Our approach uses the auxiliary concept of \textit{virtual path}. This is
discussed fully in \cite{JAROSZKIEWICZ-97A,JAROSZKIEWICZ-97B}. The basic
idea is to replace the \textit{CT }fields by functions of the node and link
variables smeared in some way over a given interval $\left[ nT,\left(
n+1\right) T\right] .$ From the point of view of \textit{CT} mechanics, this
introduces a degree of non-locality in time, but this is inevitable. For
some fields, such as the neutral scalar field, the function is a linear
interpolation, but in the case of matter fields involving gauge invariance,
such as Dirac fields, the virtual paths are highly non-linear. Fortunately,
the virtual paths for the Maxwell potential fields are straightforward. The
virtual path $\phi _{n\lambda }$ for the electric potential (a link
variable) is defined by 
\begin{equation}
\phi _{n\lambda }\left( \mathbf{x}\right) \equiv \phi _n\left( \mathbf{x}%
\right) ,\;\;\;\;0\leq \lambda \leq 1,
\end{equation}
where we use the variable $\lambda $ to interpolate between the two ends of
a given link. Given that the magnetic vector potential is a node variable,
its associated virtual path $\mathbf{A}_{n\lambda }$ is defined by 
\begin{equation}
\mathbf{A}_{n\lambda }\left( \mathbf{x}\right) \equiv \lambda \mathbf{A}%
_{n+1}\left( \mathbf{x}\right) +\bar{\lambda}\mathbf{A}_n\left( \mathbf{x}%
\right) ,\;\;\;0\leq \lambda \leq 1,
\end{equation}
where $\mathbf{A}_n$, $\mathbf{A}_{n+1}$ are the dynamically meaningful
field values at the ends of a given link and $\bar{\lambda}\equiv 1-\lambda
. $.

The above virtual paths are defined over the interval $\left[
nT,(n+1)T\right] .$ For a set of successive time intervals, we observe that
virtual paths for node variables are continuous in time, whereas virtual
paths for link variables are not. This is tied in with the nature of their
respective dynamics. Note that no physical meaning can be attributed to the
virtual path in a genuine \textit{DT} theory, and as we have stressed before 
\cite{JAROSZKIEWICZ-97A}, these paths are used only as a tool in obtaining a
system function with built in properties such as gauge invariance. Beyond
that, they have no significance.

The \textit{DT} version of Maxwell's equations comes with a \textit{DT}
version of gauge invariance. A local \textit{DT} gauge transformation
involves a \textit{gauge function }$\chi $ associated with nodes rather than
links. A gauge function value at time $n$ and position $\mathbf{x}$ will be
denoted by $\chi _n\left( \mathbf{x}\right) $ and is assumed differentiable
with respect to $\mathbf{x.}$ The virtual path for a gauge function is given
by 
\begin{equation}
\chi _{n\lambda }\left( \mathbf{x}\right) \equiv \lambda \chi _{n+1}\left( 
\mathbf{x}\right) +\bar{\lambda}\chi _n\left( \mathbf{x}\right) ,\;\;\;0\leq
\lambda \leq 1.
\end{equation}
In \textit{CT} electromagnetism a local gauge transformation is defined by
the replacements 
\begin{equation}
A^\mu \rightarrow A^{^{\prime }\mu }=A^\mu +\partial ^\mu \chi .
\end{equation}
In our theory this becomes 
\begin{eqnarray}
\phi _n &\rightarrow &\phi _n^{\prime }=\phi _n+D_n^{+}\chi _n,
\label{gauge1} \\
\mathbf{A}_n &\rightarrow &\mathbf{A}_n^{\prime }=\mathbf{A}_n-\nabla \chi
_n.  \label{gauge2}
\end{eqnarray}
When we talk about local gauge transformations, we shall mean these last two
equations.

Turning now to the physical fields, we define the gauge invariant electric
and magnetic fields via the potentials: 
\begin{equation}
\mathbf{E}_n\mathbf{\equiv -}\nabla \phi _n-D_n^{+}\mathbf{A}_n,\;\;\;\;\;%
\mathbf{B}_n\equiv \nabla \times \mathbf{A}_n.
\end{equation}
These satisfy the homogeneous \textit{DT} Maxwell equations 
\begin{equation}
\nabla \cdot \mathbf{B}_n=0,\;\;\;\;\;\nabla \times \mathbf{E}_n+D_n^{+}%
\mathbf{B}_n=\mathbf{0.}  \label{Maxwell1}
\end{equation}

To construct a gauge invariant system function, we recall that the \textit{CT%
} Lagrange density for electromagnetism is given in terms of the Faraday
tensor $F_{\mu \nu }\equiv \partial _\mu A_\nu -\partial _\nu A_\mu $, viz 
\begin{eqnarray}
\mathcal{L} &=&-\frac{_1}{^4}F_{\mu \nu }F^{\mu \nu }  \nonumber \\
&=&-\frac{_1}{^2}\partial _\mu A_\nu \partial ^\mu A^\nu +\frac{_1}{^2}%
\partial _\nu A_\mu \partial ^\mu A^\nu  \nonumber \\
&=&\frac{_1}{^2}(\dot{A}^i+\partial _i\phi )(\dot{A}^i+\partial _i\phi )+%
\frac{_1}{^2}\left( \partial _iA^j\partial _jA^i-\partial _iA^j\partial
_iA^j\right) .
\end{eqnarray}
The gauge invariant system function density for the charge free system is
obtained by replacing the fields in the above \textit{CT} Lagrange density
by their virtual path forms and integrating with respect to $\lambda $ over
the interval $\left[ 0,1\right] $; 
\begin{eqnarray}
\mathcal{F}^n &\equiv &\langle \mathcal{L}\left( \mathbf{A}_{n\lambda },\phi
_{n\lambda }\right) \rangle  \nonumber \\
&=&\langle \frac{_1}{^2}(T^{-1}\partial _\lambda A_{n\lambda }^i+\partial
_i\phi _{n\lambda })(T^{-1}\partial _\lambda A_{n\lambda }^i+\partial _i\phi
_{n\lambda })+\frac{_1}{^2}\left( \partial _iA_{n\lambda }^j\partial
_jA_{n\lambda }^i-\partial _iA_{n\lambda }^j\partial _iA_{n\lambda
}^j\right) \rangle  \nonumber \\
&=&\frac{_1}{^2}\left( D_n^{+}\mathbf{A}_n+\nabla \phi _n\right) \cdot
\left( D_n^{+}\mathbf{A}_n+\nabla \phi _n\right) +\frac{_1}{^2}\langle
\left( \partial _iA_{n\lambda }^j\partial _jA_{n\lambda }^i-\partial
_iA_{n\lambda }^j\partial _iA_{n\lambda }^j\right) \rangle ,  \label{rtf}
\end{eqnarray}
where the angular brackets denote an integral over $\lambda ,$ i.e. 
\begin{equation}
\langle f_\lambda \rangle \equiv \int_0^1f\left( \lambda \right) \,d\lambda .
\end{equation}
Here we use the virtual path replacement $\partial _t\rightarrow
T^{-1}\partial _\lambda .$ For convenience we have not multiplied the system
function by a factor $T$ as was done in earlier papers of this series, so
that it now has the physical dimensions of a Lagrangian rather than an
action.

The equations of motion for the magnetic potential (a node variable) are
given by 
\begin{equation}
\frac \partial {\partial A^i}\left\{ \mathcal{F}^n+\mathcal{F}^{n-1}\right\} 
\stackunder{c}{=}\partial _j\frac \partial {\partial _j\partial A^i}\left\{ 
\mathcal{F}^n+\mathcal{F}^{n-1}\right\} ,
\end{equation}
which reduce to 
\begin{equation}
\Box _n\mathbf{A}_n+\nabla \Lambda _n\stackunder{c}{=}\mathbf{0,}
\label{EM1}
\end{equation}
where 
\begin{equation}
\Lambda _n\equiv D_n^{-}\phi _n+\nabla \cdot S_n\mathbf{A}_n  \label{Lor}
\end{equation}
is the \textit{DT }Lorentz function. For the scalar potential (a link
variable), the equation of motion is 
\begin{equation}
\frac{\partial \mathcal{F}^n}{\partial \phi _n}\stackunder{c}{=}\nabla \cdot 
\frac{\partial \mathcal{F}^n}{\partial \nabla \phi _n},
\end{equation}
which reduces to 
\begin{equation}
D_n^{+}\nabla \mathbf{\cdot A}_n+\nabla ^2\phi _n\stackunder{c}{=}0.
\end{equation}
This implies the equation 
\begin{equation}
\Box _n\phi _n-D_n^{+}\Lambda _n\stackunder{c}{=}0.  \label{EM2}
\end{equation}
Equations $\left( \ref{EM1}-\ref{EM2}\right) $ are \textit{DT }gauge
invariant, as can be readily verified. The reason is that our system
function is gauge invariant, and this guarantees that the equations of
motion are gauge invariant.

\subsection{The \textit{DT }Lorentz gauge}

In the \textit{DT }Lorentz gauge we set 
\begin{equation}
\Lambda _n=0,
\end{equation}
and then the potentials satisfy the massless \textit{DT} Klein-Gordon
equations 
\begin{eqnarray}
&&\Box _n\mathbf{A}_n\stackunder{c}{=}\mathbf{0}, \\
&&\Box _n\phi _n\stackunder{c}{=}0.
\end{eqnarray}

It is always possible to work in this gauge, as the following argument
shows. Suppose that we started off with a configuration of fields for which
the \textit{DT} Lorentz function was non-zero, i.e. 
\begin{equation}
\Lambda _n\equiv D_n^{-}\phi _n+S_n\nabla \mathbf{\cdot A}_n\neq 0
\end{equation}
Now consider the gauge transformation 
\begin{equation}
\mathbf{A}_n^{\prime }=\mathbf{A}_n-\nabla \chi _n,\;\;\;\;\;\phi _n^{\prime
}=\phi _n+D_n^{+}\chi _n
\end{equation}
where the gauge function $\chi _n$ satisfies the \textit{DT} inhomogeneous
Klein-Gordon equation 
\begin{equation}
\Box _n\chi _n=-\Lambda _n.  \label{ss}
\end{equation}
Then we find 
\begin{equation}
\Lambda _n^{\prime }\equiv D_n^{-}\phi _n^{\prime }+S_n\nabla \mathbf{\cdot A%
}_n^{\prime }=0,
\end{equation}
as required. We note that using our experience with the \textit{DT }%
Klein-Gordon propagators in earlier papers, we may write down a particular
solution to $\left( \ref{ss}\right) $ in the form 
\begin{equation}
\chi _x=\Sigma \!\!\!\!\!\!\int_y\Delta _{Fx-y}^{}\Lambda _y,
\end{equation}
where $\Delta _{Fx}$ is the \textit{DT} scalar massless Feynman propagator
satisfying the equation 
\begin{equation}
\Box _x\Delta _{Fx}=-\delta _x.
\end{equation}

It is noteworthy that the Lorentz condition 
\begin{equation}
\Lambda _n\equiv D_n^{-}\phi _n+S_n\nabla \mathbf{\cdot A}_n=0
\end{equation}
is second order in time, that is, relates field values on three successive
nodes and on the two links between them. This will interact in some way with
the equations of motion, which are also second order, and so we can expect
trouble. We shall see that the Lorentz condition and its generalisation
demands special attention when we come to work out the electromagnetic
propagator in the \textit{DT} Feynman gauge. In particular, we will have to
consider the existence of an inverse operator, $S_n^{-1},$ which is highly
non-local in time.

\subsection{The physical fields}

Turning to the physical (gauge invariant) electric and magnetic fields, we
may write the system function density $\left( \text{\ref{rtf}}\right) $ in
the form\textbf{\ } 
\begin{equation}
\mathcal{F}^n=\frac{_1}{^2}\mathbf{E}_n^2-\frac{_1}{^6}\left( \mathbf{B}%
_{n+1}\cdot \mathbf{B}_{n+1}+\mathbf{B}_{n+1}\cdot \mathbf{B}_n+\mathbf{B}_n%
\mathbf{\cdot B}_n\right) \,,  \label{s12}
\end{equation}
which differs from the more familiar form 
\begin{equation}
\mathcal{F}^{\prime n}=\frac{_1}{^2}\mathbf{E}_n^2-\frac{_1}{^2}\mathbf{B}%
_n\cdot S_n\mathbf{B}_n
\end{equation}
by a total temporal difference and so gives the same equations of motion.
Applying Cadzow's equation to $\left( \ref{s12}\right) $ we find 
\begin{equation}
\nabla \cdot \mathbf{E}_n\stackunder{c}{=}0,\;\;\;\;\;D_n^{-}\mathbf{E}_n%
\stackunder{c}{=}\nabla \times S_n\mathbf{B}_n.  \label{Maxwell2}
\end{equation}
Using $\left( \ref{Maxwell1}\right) $ and $\left( \ref{Maxwell2}\right) $%
\textbf{\ }we find the physical electromagnetic fields satisfy the \textit{DT%
} massless Klein-Gordon equations 
\begin{equation}
\Box _n\mathbf{E}_n\stackunder{c}{=}\mathbf{0,\;\;\;}\Box _n\mathbf{B}_n%
\stackunder{c}{=}\mathbf{0.}
\end{equation}

We may readily construct the analogues of the conserved total linear
momentum and angular momentum using the method described in \textit{Appendix
B. }For example, we find the \textit{DT} analogue of the Poynting vector is 
\begin{equation}
\mathbf{P}_n\equiv \int d^3\mathbf{x}\left\{ \mathbf{E}_n\times \mathbf{B}_n+%
\frac{_1}{^6}TB_n^i\overrightarrow{\nabla }B_{n+1}^i\right\} \stackunder{c}{=%
}\mathbf{P}_{n+1}.
\end{equation}
The expression for the total angular momentum is left as a exercise.

\subsection{Maxwell's equations in the presence of charges}

In the presence of electric charges the system function density $\left( \ref
{s12}\right) $ is replaced by 
\begin{equation}
\mathcal{F}^n\left[ j\right] =\mathcal{F}^n-\phi _n\rho _n+\frac{{}_1}{{}^2}%
\mathbf{A}_{n+1}\cdot \mathbf{j}_{n+1}+\frac{{}_1}{{}^2}\mathbf{A}_n\cdot 
\mathbf{j}_n
\end{equation}
where $\rho _n\left( \mathbf{x}\right) $ and $\mathbf{j}_n\left( \mathbf{x}%
\right) $ are the discrete time charge density and charge current
respectively. The homogeneous equations $\left( \ref{Maxwell1}\right) $
remain unaltered but now the equations of motion become 
\begin{equation}
\nabla \cdot \mathbf{E}_n\stackunder{c}{=}\rho _n,\;\;\;\;\nabla \times S_n%
\mathbf{B}_n-D_n^{-}\mathbf{E}_n\stackunder{c}{=}\mathbf{j}_n.  \label{aa}
\end{equation}
These equations are consistent provided the equation of continuity 
\begin{equation}
D_n^{-}\rho _n+\nabla \cdot \mathbf{j}_n\stackunder{c}{=}0  \label{cont}
\end{equation}
for electric charge holds. The dynamical equations of motion $\left( \ref{aa}%
\right) $ may be written in the form 
\begin{eqnarray}
&&-\nabla ^2\phi _n-D_n^{+}\nabla \mathbf{\cdot A}_n\stackunder{c}{=}\rho _n.
\label{bb} \\
&&\;\;\;\;\;\;\;\Box _n\mathbf{A}_n+\nabla \Lambda _n\stackunder{c}{=}%
\mathbf{j}_n.  \label{cc}
\end{eqnarray}
Equation $\left( \ref{bb}\right) $ may also be rewritten in the form 
\begin{equation}
\Box _n\phi _n-D_n^{+}\Lambda _n\stackunder{c}{=}S_n\rho _n.  \label{dd}
\end{equation}

In the Lorentz gauge $\left( \ref{cc}\right) $ and $\left( \ref{dd}\right) $
become 
\begin{eqnarray}
&&\Box _x\phi _x\stackunder{c}{=}S_n\rho _x,  \label{class} \\
&&\Box _x\mathbf{A}_x\stackunder{c}{=}\mathbf{j}_x.
\end{eqnarray}
We note here the appearance of the non-local $T$-symmetric operator $S_n.$

Quantisation is more convenient in the Coulomb gauge, where we set 
\begin{equation}
\nabla \mathbf{\cdot A}_n=0.
\end{equation}
Then the equations of motion (\ref{bb}) and (\ref{cc}) become 
\begin{eqnarray}
&&\;\;\;\;\;\;\;\;\;\;\;\;\;\;\nabla ^2\phi _n\stackunder{c}{=}-\rho _n,
\label{AAA} \\
&&\Box _n\mathbf{A}_n+\nabla D_n^{-}\phi _n\stackunder{c}{=}\mathbf{j}_n.
\label{BBB}
\end{eqnarray}

We see once again that the scalar potential $\phi _n$ cannot be regarded as
a dynamical field in the same way as the components of the vector potential
are. Equation (\ref{AAA}) is zeroth order in time whereas (\ref{BBB}) as a
full second order dynamical equation. This is the \textit{DT} analogue of
the situation in \textit{CT }electromagnetism, where a direct application of
Dirac's constraint analysis shows that the momentum conjugate to the scalar
potential vanishes. This has important consequences when we develop our
quantisation via the Schwinger action principle, discussed next.

\subsection{comment}

It comes as a surprise to see that charge density as formulated in our
mechanics turns out to be a \textit{link} variable. A naive guess would have
us take it to be a node variable, on the grounds that electric charge is
carried by matter fields, which in our theory are node variables. This is
another example where the unravelling of dynamics from a \textit{CT} to a 
\textit{DT} framework forces us to re-evaluate our understanding of the
various components of dynamics.

For example, consider the preparation of a state containing charged
particles. In view of the above comment about charge being a link variable,
we see that it must be insufficient in some way to think of such a state as
being completely defined or specified at a given instant or node of time
only. If we wanted to measure the total charge of the system, for instance,
we would have to consider the fields on the link to which this node is
attached, and the fields on the node at the other end of this link as well.

This raises another interesting thought; the link pointing forwards in time
from a given node is different to the link pointing backwards in time.
Therefore, the meaning of what constitutes a state in \textit{DT} mechanics
must depend on whether it is regarded as an initial state or as a final
state.

It is such examples which lead to the conclusion that in \textit{DT }%
cosmology, time could not be considered to have a beginning at a point only;
it would be necessary to say something about the first link as well. If
indeed it is correct to think of our fundamental interval $T$ as equivalent
to the Planck scale $T_P$, then the origin of the universe in \textit{DT }%
mechanics would require us to regard the so-called \textit{Planck epoch} as
just the first link. Then it would be wrong in this context to imagine any
form of dynamical evolution process occurring during that epoch. This is in
direct contrast with fundamental theories such as quantum gravity and
superstring theory based on continuous time, where presumably, there is
scope for a great deal of dynamical interaction during the Planck epoch.

Carrying on this line of thought, we would have to accept that dynamical
fields on nodes such as the vector potentials $\mathbf{A}$ could only start
dynamical evolution \textit{after} the Planck epoch, since they satisfy a
second order equation of motion. On the other hand, link variables such as
the scalar potential $\phi $ would not have such a restriction. The
essential point here is that \textit{DT} mechanics alters our perception of
boundary conditions. Fortunately, our discussion in this paper involves the
present epoch, and we may assume time runs from remote past to remote future
without any qualms about boundary conditions at the origin of time.

\section{Quantisation in the Coulomb gauge}

Quantisation in \textit{DT} field theory is readily tackled via the \textit{%
DT} Schwinger action principle, which we shall now state and use to
determine the \textit{DT }electromagnetic field commutators and vacuum
functional in the Coulomb gauge. It is a merit of Schwinger's approach and
also of Feynman's path integral approach to quantisation that the emphasis
is on the physically useful amplitudes of the theory, rather than on the
operators themselves, such as occurs in the canonical quantisation process.
The process of imposing naive canonical commutators between dynamical
variables and their conjugate momenta can be expected to fail in \textit{DT}
mechanics for a number of reasons: we do not have a Hamiltonian framework in
our mechanics; we do not have a constraint theory in the fashion of Dirac
for gauge field dynamics; and the construction of conjugate momenta is
straightforward only in the case of systems which are \textit{normal }\cite
{JAROSZKIEWICZ-97A}$.$ Fortunately, none of these reasons prevent us from
quantising Maxwell's equations.

\subsection{The \textit{DT }Schwinger action principle}

When we use the \textit{DT} Schwinger action principle, we work in the
Heisenberg picture and consider matrix elements between states associated
with different times. These correspond to preparation and observation, that
is, initial and final states. If $|\Psi ,M\rangle $ is the state we have
prepared at time $MT$, and $|\Phi ,N\rangle $ is a state we are asking
questions about at time $NT>MT,$ then the infinitesimal change$\;\delta
\langle \Phi ,N|\Psi ,M\rangle $ in the transition amplitude $\langle \Phi
,N|\Psi ,M\rangle $ due to infinitesimal changes in the external sources is
defined by 
\begin{equation}
\delta \langle \Phi ,N|\Psi ,M\rangle =i\langle \Phi ,N|\delta A^{NM}|\Psi
,M\rangle ,\;\;\;N>M
\end{equation}
where $\delta A^{NM}$ is the infinitesimal change in the action sum operator 
\begin{equation}
A^{NM}\equiv T\sum_{n=M}^{N-1}\int d^3\mathbf{x}\mathcal{F}^n.
\end{equation}

In the case of the electromagnetic field, we take the free field system
function density $\left( \ref{s12}\right) $ and introduce arbitrary
infinitesimal sources $\rho _n,\;\mathbf{j}_n$ in the manner of Schwinger 
\cite{SCHWINGER}. Since these sources are arbitrary, we must take care to
ensure that the charges which couple to the electromagnetic fields satisfy
the equation of continuity $\left( \ref{cont}\right) .$ Following Schwinger
and anticipating the use of the Coulomb gauge, the system function density
in the presence of the sources is given by 
\begin{equation}
\mathcal{F}^n\left[ j\right] =\mathcal{F}^n-\phi _n\rho _n^c+\frac{{}_1}{{}^2%
}\mathbf{A}_{n+1}\cdot \mathbf{j}_{n+1}^c+\frac{{}_1}{{}^2}\mathbf{A}_n\cdot 
\mathbf{j}_n^c,  \label{ab}
\end{equation}
where 
\begin{eqnarray}
\rho _n^c\left( \mathbf{x}\right) &\equiv &\rho _n\left( \mathbf{x}\right) ,
\\
\mathbf{j}_n^c\left( \mathbf{x}\right) &\equiv &\mathbf{j}_n\left( \mathbf{x}%
\right) +\nabla _{\mathbf{x}}\Sigma \!\!\!\!\!\!\int_yG_{Cx-y}^{}\left[
D_m^{-}\rho _y+\nabla _{\mathbf{y}}\mathbf{\cdot j}_y\right]
\end{eqnarray}
are the conserved charge densities constructed out of the independent
external densities $\rho _n$ and $\mathbf{j}_n$, and $G_C$ is the Coulomb
Green's function which satisfies the equation 
\begin{equation}
\nabla _{\mathbf{x}}^2G_{Cx}=-\delta _x.
\end{equation}
This has particular solution 
\begin{equation}
G_{Cx}=\frac{\delta _n}{4\pi |\mathbf{x|}T}.
\end{equation}
We note the conserved charge densities satisfy the \textit{DT }equation of
continuity 
\begin{equation}
D_n^{-}\rho _n^c\left( \mathbf{x}\right) +\nabla _{\mathbf{x}}\cdot \mathbf{j%
}_n^c\left( \mathbf{x}\right) =0,
\end{equation}
regardless of the values of the independent densities. The above coupling
ensures that we are free to vary $\rho _n$ and $\mathbf{j}_n$ arbitrarily
whilst still ensuring that the electromagnetic fields are coupled to
conserved charges.

Functional differentiation is defined via 
\begin{equation}
\frac \delta {\delta \rho _x}\rho _y\equiv \frac \delta {\delta \rho
_n\left( \mathbf{x}\right) }\rho _m\left( \mathbf{y}\right) =\delta
_{x-y}\equiv \frac{\delta _{n-m}}T\delta ^3\left( \mathbf{x-y}\right)
\end{equation}
and similarly for the currents. Then with the external sources coupled as in 
$\left( \ref{ab}\right) $ we find

\begin{eqnarray}
\frac{i\delta }{\delta \rho _n\left( \mathbf{x}\right) }\langle \Phi
,n+1|\Psi ,n\rangle _j &=&\langle \Phi ,n+1|\phi _n\left( \mathbf{x}\right)
|\Psi ,n\rangle _j \\
\frac{-i\delta }{\delta j_n^i\left( \mathbf{x}\right) }\langle \Phi ,n|\Psi
,n-1\rangle _j &=&\frac{_1}{^2}\langle \Phi ,n|A_n^i\left( \mathbf{x}\right)
|\Psi ,n-1\rangle _j \\
\frac{-i\delta }{\delta j_n^i\left( \mathbf{x}\right) }\langle \Phi
,n+1|\Psi ,n\rangle _j &=&\frac{_1}{^2}\langle \Phi ,n+1|A_n^i\left( \mathbf{%
x}\right) |\Psi ,n\rangle _j,
\end{eqnarray}
using the Coulomb gauge (or transversality) condition 
\begin{equation}
\langle \Phi |\nabla \cdot \mathbf{A}_n\left( \mathbf{x}\right) |\Psi
\rangle =0  \label{trans}
\end{equation}
for all states $\Psi ,\,\Phi .$

If now we assume the existence of \textit{in} and \textit{out} vacua, and if 
$Z\left[ j\right] \equiv \langle 0^{out}|0^{in}\rangle _j$ denotes the
vacuum functional in the presence of the external sources, then direct
application of the \textit{DT} Schwinger action principle gives the
functional derivatives 
\begin{eqnarray}
\frac{i\delta }{\delta \rho _n\left( \mathbf{x}\right) }Z\left[ j\right]
&=&\langle 0_{out}|\phi _n\left( \mathbf{x}\right) |0_{in}\rangle _j, \\
\frac{-i\delta }{\delta j_n^i\left( \mathbf{x}\right) }Z\left[ j\right]
&=&\langle 0_{out}|A_n^i\left( \mathbf{x}\right) |0_{in}\rangle _j.
\end{eqnarray}

In the Coulomb gauge the quantum analogues of Cadzow's equations are 
\begin{eqnarray}
&&\;\;\;\;\;\;\;\;\;\;\;\nabla ^2\phi _n\left( \mathbf{x}\right) \stackunder{%
c}{=}-\rho _n^c\left( \mathbf{x}\right) , \\
&&\Box _n\langle 0^{out}|\mathbf{A}_n\left( \mathbf{x}\right) |0^{in}\rangle
_j+\nabla _{\mathbf{x}}D_n^{-}\phi _n\left( \mathbf{x}\right) Z\left[
j\right] \stackunder{c}{=}\mathbf{j}_n^c\left( \mathbf{x}\right) Z\left[
j\right] ,
\end{eqnarray}
taking the scalar potential in the Coulomb gauge to be a c-number. From
these equations we deduce 
\begin{equation}
\;\;\phi _x=\Sigma \!\!\!\!\!\!\int_yG_{Cx-y}\rho _y
\end{equation}
and 
\begin{equation}
\Box _x^{}\langle 0^{out}|A_x^i|0^{in}\rangle _j=j_x^{c\,i}Z\left[ j\right]
+\partial _i^x\partial _j^x\Sigma \!\!\!\!\!\!\int_y\mathbf{\,}%
G_{C\,x-y}j_y^{c\,j}Z\left[ j\right] .
\end{equation}
Note that we have switched notation, as we shall do frequently, using the
symbols $x,y$ to denote $\left( n,\mathbf{x}\right) $ and $\left( m,\mathbf{y%
}\right) $ respectively.

The second functional derivative of this last equation gives the \textit{DT}
time-ordered product \cite{JAROSZKIEWICZ-97A,JAROSZKIEWICZ-97C} 
\begin{equation}
\langle 0|\tilde{T}A_x^iA_y^j|0\rangle =i\Delta _{F\,x-y}^{}\delta
_{ij}+i\Sigma \!\!\!\!\!\!\int_z\Delta _{Fx-z}^{}\partial _i^z\partial
_j^zG_{C\,z-y}
\end{equation}
in the absence of the sources. This is equivalent to 
\begin{equation}
\langle 0|\tilde{T}A_n^i\left( \mathbf{x}\right) A_m^j\left( \mathbf{y}%
\right) |0\rangle =i\int \frac{d^3\mathbf{p}}{\left( 2\pi \right) ^3}\left[
\delta _{ij}-\frac{p^ip^j}{\mathbf{p}^2}\right] \tilde{\Delta}_F^{n-m}\left( 
\mathbf{p}\right) e^{-i\mathbf{p\cdot }\left( \mathbf{x-y}\right) }
\label{xd}
\end{equation}
where $\tilde{\Delta}_F^n\left( \mathbf{p}\right) $ is the fourier transform 
\begin{equation}
\tilde{\Delta}_F^n\left( \mathbf{p}\right) =\int d^{\mathbf{\ }3}\mathbf{x}%
\Delta _F^n\left( \mathbf{x}\right) e^{-i\mathbf{p\cdot x}}
\end{equation}
of the temporally indexed Greens' function, which satisfies the \textit{DT}
massless field equation 
\begin{equation}
\Box _n\Delta _F^n\left( \mathbf{x}\right) =-\frac{\delta _n}T\delta
^3\left( \mathbf{x}\right) .
\end{equation}

In the absence of external sources we may choose the radiation gauge,
defined by 
\begin{equation}
\phi _n\left( \mathbf{x}\right) =0,\;\;\;\langle \Phi |\nabla \mathbf{\cdot A%
}_n\left( \mathbf{x}\right) |\Psi \rangle =0,\;\;\;\;\forall \;\Psi ,\Phi .
\end{equation}
Using the above time ordered products $\left( \ref{xd}\right) $, we can
extract the following unequal-time commutators: 
\begin{eqnarray}
\langle 0|\left[ \tilde{A}_{n+1}^i\left( \mathbf{p}\right) ,\tilde{A}%
_n^{j+}\left( \mathbf{q}\right) \right] |0\rangle &=&-i\Gamma _p\left[
\delta _{ij}-\frac{p^ip^j}{\mathbf{p}^2}\right] \left( 2\pi \right) ^3\delta
^3\left( \mathbf{p-q}\right) \\
\langle 0|\left[ \tilde{A}_{n+1}^{i+}\left( \mathbf{p}\right) ,\tilde{A}%
_n^j\left( \mathbf{q}\right) \right] |0\rangle &=&-i\Gamma _p\left[ \delta
_{ij}-\frac{p^ip^j}{\mathbf{p}^2}\right] \left( 2\pi \right) ^3\delta
^3\left( \mathbf{p-q}\right)
\end{eqnarray}
where 
\begin{equation}
\Gamma _p=\frac{6T}{6+T^2\mathbf{p}^2}
\end{equation}
and 
\begin{eqnarray}
\tilde{A}_n^i\left( \mathbf{p}\right) &\equiv &\int d^3\mathbf{x}e^{-i%
\mathbf{p\cdot x}}A_n^i\left( \mathbf{x}\right) , \\
\tilde{A}_n^{i+}\left( \mathbf{p}\right) &\equiv &\int d^3\mathbf{x}e^{i%
\mathbf{p\cdot x}}A_n^i\left( \mathbf{x}\right) .
\end{eqnarray}
Hence we obtain the result 
\begin{equation}
\langle 0|\left[ A_{n+1}^i\left( \mathbf{x}\right) ,A_n^j\left( \mathbf{y}%
\right) \right] |0\rangle =-i\int \frac{d^3\mathbf{p}}{\left( 2\pi \right) ^3%
}e^{i\mathbf{p\cdot }\left( \mathbf{x-y}\right) }\left[ \delta _{ij}-\frac{%
p^ip^j}{\mathbf{p}^2}\right] \frac{6T}{6+T^2\mathbf{p}^2}
\end{equation}
which is precisely the same as for the scalar field discussed in \cite
{JAROSZKIEWICZ-97C} apart from the modified Kronecker delta, necessary to
preserve the transversality condition $\left( \ref{trans}\right) $.

As a final step, we may suppose that the commutators of the fields are 
\textit{c-numbers}, in the language of Dirac, and then we arrive at the
operator commutator statement 
\begin{equation}
\left[ A_{n+1}^i\left( \mathbf{x}\right) ,A_n^j\left( \mathbf{y}\right)
\right] =-i\int \frac{d^3\mathbf{p}}{\left( 2\pi \right) ^3}e^{i\mathbf{%
p\cdot }\left( \mathbf{x-y}\right) }\left[ \delta _{ij}-\frac{p^ip^j}{%
\mathbf{p}^2}\right] \frac{6T}{6+T^2\mathbf{p}^2},  \label{ee}
\end{equation}
which amounts to our \textit{DT} quantisation prescription. We may use $%
\left( \ref{ee}\right) $ and the operator equation of motion 
\begin{equation}
\Box _n\mathbf{A}_n\left( \mathbf{x}\right) =\mathbf{0}\text{, }
\end{equation}
to deduce the equal time commutators 
\begin{equation}
\left[ \tilde{A}_n^i\left( \mathbf{p}\right) ,\tilde{A}_n^{j+}\left( \mathbf{%
q}\right) \right] =0,
\end{equation}
which is equivalent to 
\begin{equation}
\left[ A_n^i\left( \mathbf{x}\right) ,A_n^j\left( \mathbf{y}\right) \right]
=0.  \label{ff}
\end{equation}

Photon creation and annihilation operators are defined by 
\begin{eqnarray}
a_n\left( \mathbf{p},\lambda \right) &\equiv &\frac i{\Gamma _p}\int d^3%
\mathbf{x\,e}^{in\theta _p-i\mathbf{p\cdot x}}\;\mathbf{\epsilon }\left( 
\mathbf{p,}\lambda \right) \mathbf{\cdot }\left[ \mathbf{A}_{n+1}\left( 
\mathbf{x}\right) -e^{i\theta _p}\mathbf{A}_n\left( \mathbf{x}\right) \right]
\\
a_n^{+}\left( \mathbf{p},\lambda \right) &\equiv &\frac{-i}{\Gamma _p}\int
d^3\mathbf{x\,e}^{-in\theta _p+i\mathbf{p\cdot x}}\;\mathbf{\epsilon \cdot }%
\left( \mathbf{p},\lambda \right) \left[ \mathbf{A}_{n+1}\left( \mathbf{x}%
\right) -e^{-i\theta _p}\mathbf{A}_n\left( \mathbf{x}\right) \right]
\end{eqnarray}
where 
\begin{equation}
\;\cos \theta _p=\frac{6-2T^2\mathbf{p}^2}{6+T^2\mathbf{p}^2}  \label{the}
\end{equation}
and 
\begin{equation}
\mathbf{\epsilon }\left( \mathbf{p},\lambda \right) \mathbf{\cdot \epsilon }%
\left( \mathbf{p},\lambda ^{\prime }\right) =\delta _{\lambda \lambda
^{\prime }},\;\;\;\;\;\;\mathbf{\epsilon }\left( \mathbf{p},\lambda \right) 
\mathbf{\cdot p=}0.
\end{equation}
In these definitions we have taken the momentum $\mathbf{p}$ to be in the
elliptic regime$.\;$From our work in previous papers with the \textit{DT}
harmonic oscillator, we know that momentum in the hyperbolic region $|%
\mathbf{p|}T>\sqrt{12}$ would give wave-functions which either decay to zero
or diverge in time. The implications are that, like scalar and Dirac
particles discussed in earlier papers of this series, there is a natural
cutoff in our mechanics in the photon spectrum. This may have important
repercussions in discussions involving for example the black body spectrum.

A direct application of the commutation rules $\left( \ref{ee}-\ref{ff}%
\right) $ gives 
\begin{eqnarray}
\left[ a_n\left( \mathbf{p},\lambda \right) ,a_n^{+}\left( \mathbf{q}%
,\lambda ^{\prime }\right) \right] &=&2|\mathbf{p|}\sqrt{1-\frac{T^2\mathbf{p%
}^2}{12}}\delta _{\lambda \lambda ^{\prime }}\left( 2\pi \right) ^3\delta
^3\left( \mathbf{p-q}\right) \\
\left[ a_n\left( \mathbf{p},\lambda \right) ,a_n\left( \mathbf{q},\lambda
^{\prime }\right) \right] &=&0,
\end{eqnarray}
which shows explicitly that we will obtain a spectrum of polarised photon
states, but only up to the parabolic barrier $|\mathbf{p|T<}\sqrt{12},$ as
discussed above. We note that an expansion in powers of $T$ gives 
\begin{equation}
\left[ a_n\left( \mathbf{p},\lambda \right) ,a_n^{+}\left( \mathbf{q}%
,\lambda ^{\prime }\right) \right] =2|\mathbf{p|}\delta _{\lambda \lambda
^{\prime }}\left( 2\pi \right) ^3\delta ^3\left( \mathbf{p-q}\right)
+O\left( T^2\right) ,
\end{equation}
which supports the results discussed in earlier papers of this series that
Lorentz symmetry emerges very rapidly from our mechanics if, as we imagine, $%
T$ is of the order of the Planck time or less.

\subsection{Comments}

i) Fixing the value of $\theta $ to equal $\theta _{\mathbf{p}}$ in $\left( 
\ref{the}\right) $ is the \textit{DT} equivalent to a mass-shell constraint
in \textit{CT }field theory;

ii) The cutoff in the high momentum photon particle spectrum occurs for a
dynamical reason. Consider the following toy model; suppose $\psi _n$ is a
complex valued variable evolving according to the \textit{DT} harmonic
oscillator equation $\cite{JAROSZKIEWICZ-97A}$ 
\begin{equation}
\psi _{n+1}\stackunder{c}{=}2\eta \psi _n-\psi _{n-1,}^{}
\end{equation}
where $\eta $ is real. Now take another complex sequence $\left( z_n\right) $
satisfying the same equation. Then we can readily prove that the
constructions 
\begin{eqnarray}
a_n &\equiv &i\left( z_n^{*}\psi _{n+1}-z_{n+1}^{*}\psi _n\right)  \nonumber
\\
a_n^{*} &=&-i\left( z_n\psi _{n+1}^{*}-z_{n+1}\psi _n^{*}\right)
\end{eqnarray}
are invariants of the motion, viz 
\begin{equation}
a_n\stackunder{c}{=}a_{n-1},\;\;a_n^{*}\stackunder{c}{=}a_{n-1}^{*},
\end{equation}
regardless of the value of $\eta $. In a field theory, the equivalent of $%
\eta $ is determined by the momentum $\mathbf{p}$.

Now we ask where the cutoff in comes from. The answer is found in the
pattern of possible behaviour found in the \textit{DT }harmonic oscillator.
We showed in \cite{JAROSZKIEWICZ-97A} that bounded motion occurs when $\eta $
is restricted to the range $-1<\eta <1$, corresponding to the trigonometric
solutions of the \textit{CT }harmonic oscillator. We call this regime the 
\textit{elliptic }region. For $\eta =\pm 1$ we have the \textit{parabolic }%
regime, and for $|\eta |>1$ we have the \textit{hyperbolic} regime, where
the solutions either diverge or collapse to zero in the limit of infinite
time. It is this which creates the cutoff. If we attempted to create
particle states with momentum in the hyperbolic regime, we would find
physically unacceptable behaviour occurring in matrix elements after a
sufficiently long time. Such states would not be stationary in the
conventional sense. An analogous phenomenon occurs in \textit{CT }quantum
wave mechanics, where we reject solutions to wave equations on the basis
that they have unacceptable behaviour at large spatial distances.

\section{The vacuum functional}

A fundamental problem in quantum gauge field theory is the construction of
the vacuum functional $Z\left[ j\right] $. In this section we shall first
solve the functional differential equations obtained via the \textit{DT }%
Schwinger action principle to obtain the pure electromagnetic vacuum
functional in the presence of external sources in the Coulomb gauge. A
functional integral approach then becomes of interest as an independent
means of calculating the same quantity and confirming that the \textit{DT}
Schwinger action principle is sound.

\subsection{The \textit{DT} Schwinger vacuum functional}

In the Coulomb gauge we take $\phi _n$ to be a classical field and the $%
\mathbf{A}_n$ to be quantum fields, satisfying the $DT$ Schwinger action
principle relations 
\begin{equation}
\phi _xZ\left[ j\right] =\frac{i\delta }{\delta \rho _x}Z\left[ j\right]
,\;\;\;\;\langle A_x^i\rangle _j=\frac{-i\delta }{\delta j_x^i}Z\left[
j\right]
\end{equation}
If now we take the equations of motion 
\begin{eqnarray}
\phi _x &=&\Sigma \!\!\!\!\!\!\int_yG_{Cx-y}\rho _y \\
\square _x\langle 0^{out}|A_x^i|0^{in}\rangle _j &=&\left( j_x^{ci}+\partial
_i^x\partial _j^x\Sigma \!\!\!\!\!\!\int_y\Delta _{Fx-y}j_y^{cj}\right)
Z\left[ j\right]
\end{eqnarray}
and write the vacuum functional as the product 
\begin{equation}
Z\left[ j\right] =Z\left[ \rho \right] Z\left[ \mathbf{j}\right]
\end{equation}
then we may readily integrate the resulting functional differential
equations to find the \textit{DT }Schwinger action functional 
\begin{eqnarray}
Z\left[ j\right] &\sim &\exp \left\{ -\frac{_1}{^2}i\Sigma
\!\!\!\!\!\!\int_x\Sigma \!\!\!\!\!\!\int_y\mathbf{\,}\left[ \rho
_xG_{Cx-y}\rho _y+\mathbf{\,}j_x^i\Delta _{Fx-y}j_y^i-\Sigma
\!\!\!\!\!\!\int_z\nabla \mathbf{\cdot j}_x\Delta _{Fx-y}G_{Cy-z}\nabla
\cdot \mathbf{j}_z\right] \right\} .  \nonumber \\
&&  \label{Sch}
\end{eqnarray}

\subsection{ The \textit{DT} Faddeev-Popov vacuum functional}

It is a remarkable feature of quantum gauge field dynamics that we may use
two completely different routes to determine the \textit{CT} vacuum
functional $Z\left[ j\right] $ expressed as a path integral in \textit{CT}
field theory. Superficially these seem quite different. One way is to work
out the details of the physical phase space dynamics via Dirac's constraint
mechanics and then define $Z\left[ j\right] $ in terms of the non-redundant
or physical degrees of freedom. The other route is via the Faddeev-Popov
gauge symmetry approach. These apparently unrelated approaches give the same
result.

The former approach involves the Hamiltonian \cite{ABERS_&_LEE,RAMOND} and
therefore has no analogue in \textit{DT} mechanics as far as we understand
at this time (because we do not have a Hamiltonian approach in our
mechanics, there being no generator of continuous translations in discrete
time).

Fortunately, the Faddeev-Popov method approach uses gauge symmetry arguments
to achieve the same end, and there is nothing in those arguments which
prevents us from employing exactly the same method in \textit{DT} gauge
theory. By following the standard arguments \cite{ABERS_&_LEE,RAMOND} we
arrive at the expression 
\begin{equation}
Z\left[ j\right] \sim \prod_n\int \left[ d\mathbf{A}_n\right] \left[ d\phi
_n\right] \Delta _g\left[ \mathbf{A,}\phi \right] \delta \left[ g\left( 
\mathbf{A,}\phi \right) \right] \exp \left\{ iS\left[ j\right] \right\} \,
\label{F-P}
\end{equation}
where $g$ is the gauge fixing function. Here we have taken the gauge fixing
function to reside on the nodes. This is consistent with the Coulomb gauge
discussed next and with the \textit{DT }analogues of the Landau and Feynman
gauges discussed after that.

If we choose the Coulomb gauge then 
\begin{equation}
g_n=\nabla \cdot \mathbf{A}_n,
\end{equation}
the $\Delta _g\;$factor is independent of the fields, and the action
integral reduces to 
\begin{equation}
S\left[ j\right] =\Sigma \!\!\!\!\!\!\int_x\left\{ -\frac{_1}{^2}\phi \nabla
^2\phi -\frac{_1}{^2}\mathbf{A}\cdot \square \mathbf{A}-\rho \phi +\mathbf{j}%
\cdot \mathbf{A}\right\} .
\end{equation}
Then using the functional integral result 
\begin{equation}
\int \left[ d\psi \right] \exp \left\{ i\int dx\,\,\psi M\psi +i\int dxj\psi
\right\} \sim \sqrt{\det M^{-1}}\exp \left\{ -\frac{_1}{^2}i\int
dx\,jM^{-1}j\right\}
\end{equation}
we recover the Schwinger vacuum functional $\left( \text{\ref{Sch}}\right) $
exactly. This confirms the consistency of our approach.

\section{More general gauges}

The Faddeev-Popov approach allows us to consider more general gauges. We now
follow the standard gauge fixing approach to construct the \textit{DT }%
analogues of the Landau and Feynman gauge propagators. First we note that
the vacuum functional $\left( \ref{F-P}\right) $ is independent of the gauge
fixing function, so we can choose 
\begin{equation}
g_x\equiv \Lambda _x-\omega _x,
\end{equation}
where $\omega _x\equiv \omega _n\left( \mathbf{x}\right) $ is an arbitrary
function on \textit{DT }space-time and $\Lambda $ is the \textit{DT} Lorentz
function $\left( \ref{Lor}\right) $. Now we can functionally integrate over $%
\omega $, weighting the integrand by a suitable weighting function. The
essential step here is to choose the weighting function to be the
exponential 
\begin{equation}
\exp \left( \frac{-i}{2\alpha }\Sigma \!\!\!\!\!\!\int_x\omega _x%
\overrightarrow{N\left( U_n\right) }\omega _x\right)
\end{equation}
where $N\left( U_n\right) $ is some $T-$symmetric operator to be determined
and $\alpha $ is a constant. After integrating out $\omega $ via the
functional delta function, we can take the vacuum functional to be given by 
\begin{equation}
Z\left[ j\right] \sim \int \left[ d\mathbf{A}\right] \left[ d\phi \right]
\exp \left\{ iS_0+iS_G+iS_j\right\} ,
\end{equation}
where $S_0$, $S_G$ and $S_j$ are the free electromagnetic action sum, the
gauge fixing term and the source term respectively. We now consider these
three terms separately. The most efficient approach is to work with the four
component object 
\begin{equation}
A_n^\mu \equiv \left( \phi _n,\mathbf{A}_n\right) ,
\end{equation}
which looks like a Lorentz four-vector, but it should be stressed that this
is a matter of convenience only. For one thing, $A_n^0\equiv \phi _n$ is a
link variable whereas the other components are node variables.

Suppressing the space-time indices $\left( n,\mathbf{x}\right) ,$ we find 
\begin{eqnarray}
S_0 &\equiv &\int \!\!\!\!\!\!\Sigma \left\{ -\frac{_1}{^2}A^i\square A^i+%
\frac{_1}{^2}\partial _iA^iS\partial _jA^j-\frac{_1}{^2}\phi \nabla ^2\phi
+\partial _i\phi D^{+}A^i\right\}  \nonumber \\
&=&\frac{_1}{^2}\int \!\!\!\!\!\!\Sigma A^\mu R_{\mu \nu }A^\nu
\end{eqnarray}
where the operator 
\begin{equation}
R_{\mu \nu }\equiv \left[ 
\begin{tabular}{cc}
$-\nabla ^2$ & $-D^{+}\partial _i$ \\ 
$-D^{-}\partial _i$ & $-\square \delta _{ij}-S\partial _i\partial _j$%
\end{tabular}
\right] .
\end{equation}
written in $(1,3)\times (1,3)$ block form acts to the right.

The gauge fixing term is given by 
\begin{eqnarray}
S_G &\equiv &-\frac 1{2\alpha }\int \!\!\!\!\!\!\Sigma \left( D^{-}\phi
+S\partial _iA^i\right) N\left( D^{-}\phi +S\partial _iA^i\right)  \nonumber
\\
&=&+\frac{_1}{^2}\alpha ^{-1}\int \!\!\!\!\!\!\Sigma A^\mu S_{\mu v}A^\nu
\end{eqnarray}
where the operator 
\begin{equation}
S_{\mu \nu }=\left[ 
\begin{tabular}{cc}
$ND^2$ & $ND^{+}S\partial _i$ \\ 
$NSD^{-}\partial _i$ & $NS^2\partial _i\partial _j$%
\end{tabular}
\right]
\end{equation}
acts to the right. The current term is just 
\begin{equation}
S_j\equiv -\int \!\!\!\!\!\!\Sigma j_\mu A^\mu ,
\end{equation}
where $j_\mu \equiv \left( \rho ,-\mathbf{j}\right) .$ Hence 
\begin{equation}
Z\left[ j\right] \sim \int \left[ dA^\mu \right] \exp \left\{ \frac{_1}{^2}%
i\int \!\!\!\!\!\!\Sigma A^\mu M_{\mu \nu }A^\nu -i\int \!\!\!\!\!\!\Sigma
j_\mu A^\mu \right\}
\end{equation}
where 
\begin{eqnarray}
M_{\mu \nu } &\equiv &R_{\mu \nu }+\alpha ^{-1}S_{\mu \nu }  \nonumber \\
&=&\left[ 
\begin{tabular}{cc}
$-\nabla ^2+\alpha ^{-1}ND^2$ & $\left( \alpha ^{-1}NS-1\right)
D^{+}\partial _i$ \\ 
$\left( \alpha ^{-1}NS-1\right) D^{-}\partial _i$ & $-\square \delta
_{ij}+\left( \alpha ^{-1}NS^2-S\right) \partial _i\partial _j$%
\end{tabular}
\right]
\end{eqnarray}
Assuming $M_{\mu \nu }$ has an inverse, we may integrate to find 
\begin{equation}
Z\left[ j\right] \sim \exp \left\{ \frac{_i}{^2}j_\mu G^{\mu \nu }j_\nu
\right\}
\end{equation}
where 
\begin{equation}
M_{\mu \nu }G^{\nu \lambda }=-\delta _\mu ^\lambda .  \label{ewq}
\end{equation}
The calculation of $G^{\nu \lambda }$ goes as follows. We note $\left( \ref
{ewq}\right) $ is equivalent to 
\begin{equation}
\Sigma \!\!\!\!\!\!\int_y\overrightarrow{M_{\mu \nu x}}\delta
_{x-y}G_{y-z}^{\nu \lambda }=-\delta _\mu ^\lambda \delta _{x-z},
\end{equation}
i.e 
\begin{equation}
\Sigma \!\!\!\!\!\!\int_xe_{px}\overrightarrow{M_{\mu \nu x}}G_x^{\nu
\lambda }=-\delta _\mu ^\lambda .
\end{equation}
Assuming $G_x^{\nu \lambda }$ has a \textit{DT }fourier transform $\tilde{G}%
_q^{\nu \lambda }$, we may write 
\begin{equation}
\Sigma \!\!\!\!\!\!\int_xe_{px}\overrightarrow{M_{\mu \nu x}}\oint_q\bar{e}%
_{qx}\tilde{G}_q^{\nu \lambda }=-\delta _\mu ^\lambda .
\end{equation}
Now $\bar{e}_{qx}$ is an eigenfunction of the operator $\overrightarrow{%
M_{\mu \nu x}},$ so we may write 
\begin{equation}
\overrightarrow{M_{\mu \nu x}}\bar{e}_{qx}=\tilde{M}_{\mu \nu q}\bar{e}_{qx}
\end{equation}
where 
\begin{equation}
\tilde{M}_{\mu \nu q}\equiv \left[ 
\begin{tabular}{cc}
$\mathbf{q}^2+\alpha ^{-1}NuD^2u$ & $i\left( 1-\alpha ^{-1}NuSu\right)
D^{-}uq^i$ \\ 
$i\left( 1-\alpha ^{-1}NuSu\right) D^{+}uq^i$ & $q^2\delta _{ij}-\left(
\alpha ^{-1}NuS^2u-Su\right) q^iq^j$%
\end{tabular}
\right]  \label{mat}
\end{equation}
in block form, where $q\equiv \left( u,\mathbf{q}\right) $. Hence we find 
\begin{equation}
\tilde{M}_{\mu \nu p}\tilde{G}_p^{\nu \lambda }=-\delta _\mu ^\lambda .
\end{equation}
Now if we write 
\begin{eqnarray}
\tilde{M}_{\mu \nu p} &\equiv &\left[ 
\begin{tabular}{cc}
$a$ & $b\mathbf{p}^T$ \\ 
$b^{*}\mathbf{p}$ & $c+d\mathbf{pp}^T$%
\end{tabular}
\right]  \nonumber \\
\tilde{G}_p^{\nu \lambda } &\equiv &\left[ 
\begin{tabular}{cc}
$A$ & $B^{*}\mathbf{p}^T$ \\ 
$B\mathbf{p}$ & $C+D\mathbf{pp}^T$%
\end{tabular}
\right]
\end{eqnarray}
then a solution is 
\begin{eqnarray}
A &=&-\frac{c+d\mathbf{p}^2}{ac+\mathbf{p}^2\Delta },\;\;\;\;\;B=\frac{b^{*}%
}{ac+\mathbf{p}^2\Delta }  \nonumber \\
C &=&\frac 1c,\;\;\;\;\;\;\;\;\;\;\;\;\;\;\;D=\frac \Delta {c\left[ ac+%
\mathbf{p}^2\Delta \right] }
\end{eqnarray}
where $\Delta \equiv ad-b^{*}b.$ Now from $\left( \ref{mat}\right) $ we read
off 
\begin{eqnarray}
a &=&\mathbf{p}^2+\alpha ^{-1}NzD^2z,\;\;\;\;\;\;\;b=i\left( 1-\alpha
^{-1}NzSz\right) D^{-1}z  \nonumber \\
c &=&p^2\;\;\;\;\;\;\;\;\;\;\;\;\;\;\;\;\;\;\;d=Sz-\alpha ^{-1}NzS^2z
\end{eqnarray}
and now we may readily compute the components of $\tilde{G}_p^{\nu \lambda
}. $

The solution is complicated in the case of arbitrary $N$. We may now make a
careful choice which reduces the complexity and brings us closer to the 
\textit{CT} results. The particular choice which works here is 
\begin{equation}
Nz\equiv \frac 1{Sz},
\end{equation}
which means that $N\left( U_n\right) $ is the operator inverse of $S\left(
U_n\right) .$

Some concern may be expressed at this choice, as it involves step operators
of all orders and may be undefined for some sequences. However, we can allay
fears by noting that in the \textit{DT }fourier transform space, the
reciprocal of\textit{\ }$Sz\equiv \left( z+4+1/z\right) /6$ exists, provided
we are on the unit circle in the complex $z$ plane. We may use this to 
\textit{define} the inverse operator $N\left( U_n\right) =S^{-1}\left(
U_n\right) $ by its action in \textit{DT} fourier transform space, viz 
\begin{equation}
S^{-1}\left( U_n\right) f_n\equiv \oint_z\frac{\tilde{f}_z}{z^nSz}.
\end{equation}
We note that $Sz$ has two zeros, located at $z=-2+\sqrt{3}$ and $z=-2-\sqrt{3%
}$. The first is inside the unit circle and contributes an additional simple
pole away from the origin. Hence the above integral should be well defined
and readily evaluated using the calculus of residues.

With the above choice for $Nz$, the \textit{DT }Landau gauge propagator is
obtained by setting $\alpha =0$ and then we find 
\begin{equation}
\tilde{G}_L^{\mu v}\left( p\right) =\frac 1{p^2p^2}\left[ 
\begin{tabular}{cc}
$-S^2z\mathbf{p}^2$ & $iSzD^{-}z\mathbf{p}^T$ \\ 
$iSzD^{+}z\mathbf{p}$ & $-p^2\delta _{ij}-Szp^ip^j$%
\end{tabular}
\right]
\end{equation}
in block form. This tends to the correct \textit{CT} Landau gauge propagator 
\begin{eqnarray}
\tilde{G}_L^{\mu v}\left( p\right) &=&\frac 1{p^2}\left( \eta ^{\mu \nu }-%
\frac{p^\mu p^\nu }{p^2}\right)  \nonumber \\
&=&\frac 1{p^2p^2}\left[ 
\begin{tabular}{cc}
$-\mathbf{p}^2$ & $-p^0\mathbf{p}^T$ \\ 
$-p^0\mathbf{p}$ & $-p^2\delta _{ij}-p^ip^j$%
\end{tabular}
\right]
\end{eqnarray}
in the \textit{CT} limit 
\begin{equation}
T\rightarrow 0,\;\;Sz\rightarrow 1,\;\;\;-iD^{+}z\rightarrow
p^0,\;\;\;\;\;-iD^{-}z\rightarrow p^0.
\end{equation}
Here we have used the reparametrisation 
\begin{equation}
z\equiv e^{i\theta }
\end{equation}
where 
\begin{equation}
\theta =Tp^0+O\left( T^3\right) ,
\end{equation}
as discussed in \cite{JAROSZKIEWICZ-97C}.

The \textit{DT} Feynman gauge propagator is obtained by setting $\alpha =1$
and choosing $Nz=\left( Sz\right) ^{-1}$ gives 
\begin{equation}
\tilde{G}_F^{\mu v}\left( p\right) =\frac 1{p^2}\left[ 
\begin{tabular}{cc}
$Sz$ & $\mathbf{0}^T$ \\ 
$\mathbf{0}$ & $-\delta _{ij}$%
\end{tabular}
\right] ,  \label{DTF}
\end{equation}
which tends to the \textit{CT }Feynman gauge propagator 
\begin{equation}
\tilde{G}_F^{\mu v}\left( p\right) =\frac{\eta ^{\mu \nu }}{p^2}=\frac
1{p^2}\left[ 
\begin{tabular}{cc}
$1$ & $\mathbf{0}^T$ \\ 
$\mathbf{0}$ & $-\delta _{ij}$%
\end{tabular}
\right]
\end{equation}
in the $CT$ limit, as expected.

Two comments on this last result are in order.

\begin{enumerate}
\item  Our theory has not assumed any metric structure to space-time. We
note the appearance of a factor $Sz$ in the $0-0$ component in $\left( \ref
{DTF}\right) .$ The precise relationship of our \textit{DT} space-time and
any \textit{DT} analogue of the \textit{CT }metric tensor in special and
general relativity awaits further investigation;

\item  In the \textit{DT }Feynman gauge we find the first functional
derivatives with respect to the external sources of the vacuum functional
gives 
\begin{equation}
\langle 0_{out}|\phi _x|0_{in}\rangle _j=\frac{i\delta }{\delta \rho _x}%
Z\left[ j\right] =S_n\rho _nZ\left[ j\right] ,
\end{equation}
which is consistent with the classical equation $\left( \ref{class}\right) .$
\end{enumerate}

\section{Concluding remarks}

The stage is now set for the next paper in this series, which will be to
develop the \textit{DT }Feynman rules for \textit{QED} and investigate the
renormalization programme. One of the principal motivations of this series
was the possibility that the introduction of a new scale $T$ would allow a
more convincing approach to the elimination of divergences than other
formulations. Whether this works remains to be seen. We note that one of the
early and successful regularisation methods, the Pauli-Villiars technique,
employs the introduction of a very large mass scale. In some sense, this is
the other side of the coin to our approach, which is to consider very small
time scales. Either way, there is a suggestion or hint that something
intimately involved with cosmological scales, either extremely large or
extremely small, is at the core of the problems with field theory.

\section{Acknowledgments}

One of us (\textit{KN) }thanks the Crowther Fund of the Open University for
financial support during the course of this work.

\appendix

\section{A first order formulation}

In the first paper of this series \cite{JAROSZKIEWICZ-97A} we interpreted
Cadzow's equations of motion $\left( \ref{Cad}\right) $ as an expression of
momentum conservation at the node sites. The momentum in that context is
clearly a node quantity. We shall call such a momentum a \textit{node
momentum. }In this section we consider the possibility of introducing link
variables related to node variables via Legendre transformations and these
we shall call \textit{link momenta. }The node and link momentum concepts are
different in \textit{DT} mechanics, but it turns out that in the \textit{CT}
limit $T\rightarrow 0$ they coincide. It is part of the problem of
quantisation in \textit{DT} mechanics that it is not clear \textit{a priori}
whether node or link momenta should be used in setting up canonical
quantisation algebras such as the Weyl-Heisenberg algebra. In the previous
papers of this series we avoided this question by using the \textit{DT}
analogue of the Schwinger action principle to obtain quantum commutation
relations consistent with our \textit{DT} mechanics.

The reason we should consider link momenta is because there is more than one
way to arrive at the Euler-Lagrange equations for configuration space
variables in \textit{CT} mechanics. The usual way is called the second order
formulation, wherein we start with the Lagrangian $L\equiv L\left( \mathbf{q,%
\dot{q},}t\right) $ and use Hamilton's principle directly to get the
Euler-Lagrange equations 
\begin{equation}
\frac d{dt}\left( \frac{\partial L}{\partial \mathbf{\dot{q}}}\right) 
\stackunder{c}{=}\frac{\partial L}{\partial \mathbf{q}}.
\end{equation}

Another route is via the so-called first order formulation, which is related
to the Hamiltonian phase space approach, but pretends to live in an extended
configuration space. We define the conjugate momenta 
\begin{equation}
\mathbf{p\equiv }\frac{\partial L}{\partial \mathbf{\dot{q}}},
\end{equation}
construct the Hamiltonian $H\left( \mathbf{p,q,}t\right) ,$ and then define
the first order Lagrangian 
\begin{equation}
\tilde{L}\equiv \mathbf{p\cdot \dot{q}-}H.
\end{equation}
Now we apply Hamilton's action principle to $\tilde{L}$ and get the extended
configuration space equations 
\begin{eqnarray}
&&\frac d{dt}\left( \frac{\partial \tilde{L}}{\partial \mathbf{\dot{q}}}%
\right) -\frac{\partial \tilde{L}}{\partial \mathbf{q}}\stackunder{c}{=}%
\mathbf{0,}  \label{one} \\
&&\;\;\;\;\;\;\;\;\;\;\;\;\;\,\frac{\partial \tilde{L}}{\partial \mathbf{p}}%
\stackunder{c}{=}\mathbf{0.}  \label{two}
\end{eqnarray}
These equations are disguised versions of Hamilton's equations

\begin{equation}
\mathbf{\dot{q}}\stackunder{c}{=}\frac{\partial H}{\partial \mathbf{p}}%
,\;\;\;\;\;\mathbf{\dot{p}}\stackunder{c}{=}-\frac{\partial H}{\partial 
\mathbf{q}}.
\end{equation}
in phase space and are equivalent to the \textit{CT} Euler-Lagrange
equations obtained from the second order formulation. This approach to
configuration space mechanics is the basis of the ADM formulation of quantum
gravity \cite{ADM-1}. We note that equations (\ref{one}) and (\ref{two}) are
remarkably similar to the two sorts of \textit{DT }equations of motion (\ref
{Cad}) and (\ref{Cad2}) for node and link variables respectively. This leads
us to expect the \textit{DT }analogue of phase space momentum to be a link
variable, rather than a node variable.

When it comes to gauge theories, the construction of link momentum for
matter fields is altered in a subtle way related to the virtual paths used
to maintain gauge invariance, but none of this subtlety is seen or required
in the \textit{CT} limit.

For completeness, we show that we may rewrite the free electromagnetic
equations in a first order formulation. We introduce the gauge invariant
link variable $\mathbf{\pi }_n$ and define the system function 
\begin{equation}
\mathcal{F}_1^n\equiv \mathbf{\pi }_n\cdot D_n^{+}\mathbf{A}_n-\frac{{}_1}{%
{}^2}\mathbf{\pi }_n\cdot \mathbf{\pi }_n+\mathbf{\pi }_n\cdot \nabla \phi
_n-\frac{_1}{^4}\langle \mathbf{B}_{n\lambda }\cdot \mathbf{B}_{n\lambda
}\rangle .
\end{equation}
Then the equations of motion are 
\begin{eqnarray}
&&D_n^{-}\mathbf{\pi }_n+S_n\nabla \times \mathbf{B}_n\stackunder{c}{=}%
\mathbf{0,} \\
&&\;\;\;\;\;\;\;\;\;\;\;\;\;\nabla \cdot \mathbf{\pi }_n\stackunder{c}{=}0,
\\
&&\;\;\;\;\;\;\;\;\;\;\;\;\;\;\;\;\;\mathbf{\pi }_n\stackunder{c}{=}D_n^{+}%
\mathbf{A}_n+\nabla \phi _n,
\end{eqnarray}
which are equivalent to the second order equations derived above. By
inspection, we see that $\mathbf{\pi }_n=-\mathbf{E}_n,$which was to be
expected from an analysis of the \textit{CT} first order formulation.

\section{Invariants of the motion}

Consider a symmetry of the system function, i.e. a transformation of the
node and link variables 
\begin{equation}
A_n^\alpha \rightarrow A_n^\alpha +\delta A_n^\alpha ,\;\;\;\phi _n^\alpha
\rightarrow \phi _n^\alpha +\delta \phi _n^\alpha
\end{equation}
such that the system function is left unchanged. Then 
\begin{eqnarray}
\delta F^n &=&\int d^3\mathbf{x}\left\{ \delta A_n^\alpha \left( \mathbf{x}%
\right) \frac \delta {\delta A_n^\alpha \left( \mathbf{x}\right) }F^n+\delta
A_{n+1}^\alpha \left( \mathbf{x}\right) \frac \delta {\delta A_{n+1}^\alpha
\left( \mathbf{x}\right) }F^n\right.  \nonumber \\
&&\;\;\;\;\;\;\;\;\;\;\;\;\;\;\;\;\;\;\;\;\;\left. +\sum_{n=M}^{N-1}\delta
\phi _n\left( \mathbf{x}\right) \frac \delta {\delta \phi _n\left( \mathbf{x}%
\right) }F^n\right\} =0
\end{eqnarray}
Using the \textit{DT }equations of motion $\left( \ref{Cad},\ref{Cad2}%
\right) $ we find the invariant of the motion 
\begin{eqnarray}
C^n &\equiv &T\int d^3\mathbf{x}\delta A_n^\alpha \left( \mathbf{x}\right)
\frac \delta {\delta A_n^\alpha \left( \mathbf{x}\right) }F^n  \nonumber \\
&=&T\int d^3\mathbf{x}\delta A_n^\alpha \left( \mathbf{x}\right) \left\{ 
\frac{\partial \mathcal{F}^n}{\partial A_n^\alpha \left( \mathbf{x}\right) }%
-\nabla \cdot \frac{\partial \mathcal{F}^n}{\partial A_n^\alpha \left( 
\mathbf{x}\right) }\right\} \stackunder{c}{=}C^{n+1},
\end{eqnarray}
which we refer to as a Maeda-Noether invariant \cite{MAEDA-81}. Although
variations of the link variables are involved in the overall variation (and
indeed are essential), partial derivatives with respect to link variables do
not occur in the explicit construction of the above invariant. This
underlines the fact that link variables are not equivalent to node variables
in a dynamical sense.

\pagebreak

\end{document}